\documentclass[a4paper,11pt]{article}
\pdfoutput=1 

\usepackage{jcappub} 

\usepackage{orcidlink}

\usepackage{newtxtext,newtxmath}
\usepackage{booktabs}
\usepackage{tabularx}
\usepackage{placeins}
\usepackage{etoolbox}

\usepackage[T1]{fontenc}
\usepackage{xcolor}

\definecolor{myblue}{RGB}{0,70,140}        

\definecolor{darkblue}{RGB}{0,51,102}

\usepackage{titlesec}
\titleformat{\paragraph}[runin]
  {\normalfont\itshape\mdseries}
  {\theparagraph}
  {1em}
  {}

\usepackage{graphicx}	
\usepackage{amsmath}	

\newcommand*\diff{\mathop{}\!\mathrm{d}}
\newcommand*\Leg{\mathsf{L}}
\makeatletter
\newcommand*\eqalias[1]{{\protected@edef\@currentlabel{\theequation}\label{#1}}}
\makeatother

\title{A statistical framework for the modelling of synchrotron emission}

\author{Zheng Zhang\note{Corresponding author.}\orcidlink{0000-0002-9154-2803}\textsuperscript{1},}
\author{Jens Chluba}
\author{and Robert Watson}
\affiliation{Jodrell Bank Centre for Astrophysics, University of Manchester, \\
Manchester, M13 9PL, United Kingdom}

\emailAdd{Zheng.Zhang@Manchester.ac.uk}
\emailAdd{Jens.Chluba@Manchester.ac.uk}

\abstract{
Flexible and fast synchrotron models are needed to interpret and separate emission components across radio and cosmic microwave background (CMB) frequencies. We develop a statistical framework for predicting and fitting optically thin synchrotron emission with Faraday rotation. It approximates spectra as linear combinations of basis spectra derived from single-electron radiation, weighted by generalised moments of the electron population, magnetic field and Faraday depth, including the correlations between them. The moments can be fitted as linear unknowns or linked through a specified population model. Tests on power-law, curved and aged electron populations, including correlated ensembles, demonstrate accurate predictions and representations of intensity and linear polarisation. Retaining correlations through joint moments avoids biases that arise when assuming statistical independence. At low noise, the intensity fits also outperform the tested empirical spectral laws for spectra that depart from a curved power law.
Extrapolating lower-frequency fits to CMB frequencies, we find that the in-band spectral shape does not by itself fix the high-frequency departure from a power law, because for the examples used here the high-energy electrons that dominate at CMB frequencies are only weakly constrained by lower-frequency data. With broad frequency coverage and low noise, fits without an assumed electron spectrum extrapolate the tested smooth spectra to within a few per cent. An electron-spectrum model that contains the true spectrum reaches this accuracy from narrower bands and noisier data, while an incorrectly chosen model can fit the band within the noise but still miss the extrapolated intensity by tens of per cent.
Because the moments are explicitly defined by the distribution, the fits also constrain the underlying physical parameters. The framework is implemented in the companion package \textsc{SyncEM}.
}


\begin{document}
\maketitle
\flushbottom



\section{Introduction}
\label{sec: introduction}

Diffuse Galactic synchrotron radiation is generated by cosmic-ray electrons spiralling in the Galactic magnetic field. It dominates much of the radio sky and remains an important contaminant at low microwave frequencies~\cite{planck2016foregrounds}. Its polarised emission furthermore is an important foreground for  primordial CMB $B$-modes, with low-frequency observations helping to constrain its contribution in the CMB observing bands~\cite{Irfan2015, bicep2021constraints}. Uncertainty in its intensity spectrum can also degrade constraints on CMB spectral distortions~\cite{abitbol2017distortions, Rotti2021SD, Sabyr2025y, Coulon2026foregrounds} and complicates observations of the redshifted 21-cm signal~\cite{pritchard2012cosmology}. The required modelling accuracy depends on the observable, observing band, and sky region.

Synchrotron spectral energy distributions (SEDs) are modelled at two levels. Empirical spectral laws, such as power laws with curvature terms, are cheap to evaluate and fit, and \textsc{PySM} uses them to extrapolate sky maps~\cite{thorne2017pysm,zonca2021pysm3,pegsg2025pysm}. Alongside these empirical descriptions of the spectral shape~\cite{kogut2012synchrotron}, physical forward models such as \textsc{GALPROP}~\cite{strong2000diffuse,orlando2013synchrotron}, \textsc{hammurabi X}~\cite{wang2020hammurabi} and \textsc{synax}~\cite{synax2024} calculate emission from specified cosmic-ray and magnetic-field distributions, while \textsc{GMOSS} combines physical components in a sky-spectrum model~\cite{rao2017gmoss}. Assuming a particular parametrisation limits the populations and signals a model can represent, while nonlinear SED fitting can be computationally expensive. We therefore seek a linear spectral representation that supports efficient SED fitting, uses basis functions derived from the emission physics and in the most general case does not require prescribed forms for the underlying parameter distributions.

Such a representation can be constructed through moment expansions of the emission model. Expanding each component's spectrum about reference (or pivot) parameters gives basis spectra whose coefficients are population averages of parameter products, including products of different parameters. For synchrotron emission, previous work has applied this construction to spectral-law parameters~\cite{chluba2017rethinking,vacher2022moment,vacher2023spin}. We extend this construction to the physical synchrotron emission model, deriving basis functions from the radiation physics and linking their coefficients to the emitting population, following recent work on spinning dust~\cite{zhangSpectralSignaturesSpinning2026,zhangSpyDustImprovedExtended2025}.

\begin{figure}[tbp]
\centering
\includegraphics[width=\textwidth]{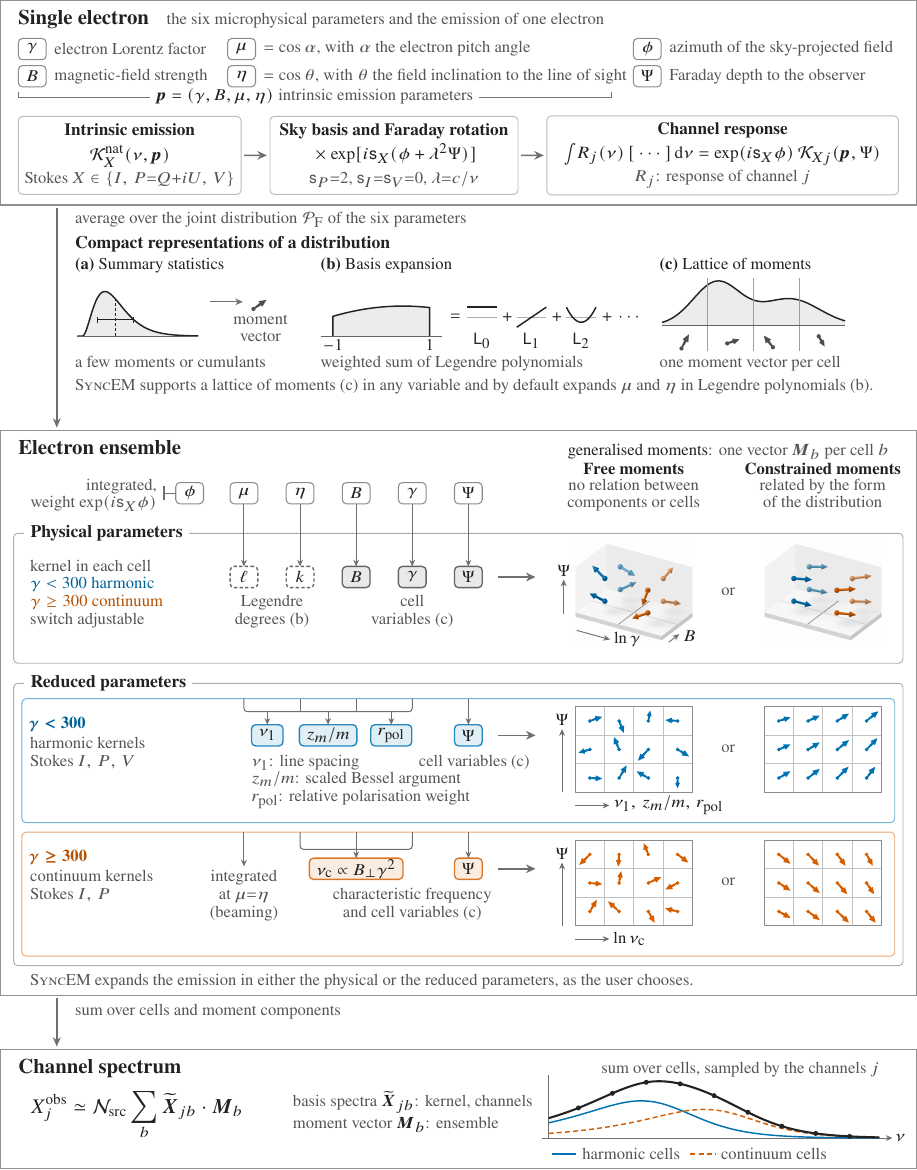}
\\
\caption{Overview of the framework. Single-electron emission and the channel responses fix the basis spectra, and the ensemble enters through one moment vector per cell. Sections~\ref{sec: single particle}--\ref{sec: moment expansion} define the symbols, and Eq.~\eqref{eq: harmonic shape combinations} defines $\nu_1$, $z_m/m$ and $r_{\rm pol}$.}
\label{fig: framework overview}
\end{figure}

We consider optically thin synchrotron emission, generally valid above the synchrotron self-absorption frequency \cite{rybicki1979radiative}, together with internal Faraday rotation \cite{burn1966depolarization}, allowing emission and rotation to occur within the same medium.
Starting from the single-electron harmonic Stokes emission \cite{leung2011numerical,pandya2016symphony}, we apply Faraday rotation and integrate over the observing channels to obtain each electron's contribution to the measured spectrum (Fig.~\ref{fig: framework overview}, top). Above a chosen electron energy, the classical continuum kernels~\cite{westfold1959polarization,rybicki1979radiative} replace the harmonic kernels and give the intensity and linear polarisation at lower cost. The ensemble SED follows by averaging these contributions over the joint distribution of electron energies, pitch angles, magnetic-field strengths and orientations, and Faraday depths.

To make the SED synthesis tractable for general parameter distributions, we combine three compact representations of this distribution. For an independent Faraday screen, a few cumulants summarise the Faraday depths. The electron pitch-angle and magnetic-field inclination distributions are expanded in Legendre polynomials of their angular cosines. For parameters spanning a broad range, such as electron energy, we divide the range into cells and describe each cell's contribution through a vector of local moments. We build the spectral basis either in the physical parameters or in fewer combinations of them that set the spectral shape. Each channel spectrum is then a linear combination of basis spectra weighted by the moments (Fig.~\ref{fig: framework overview}, bottom).

The resulting linear spectral basis supports both forward predictions from specified populations and linear SED fits with the moments as unknowns. The companion package \textsc{SyncEM} implements these synthesis and fitting procedures in \textsc{jax}~\cite{bradbury2018jax} (see Code availability). In the SED fitting problem, we can incorporate different levels of prior knowledge. Without imposing a known form of distribution, the moments can be treated as {\it free} variables, with no prescribed relations linking different components and cells. We call this treatment ``free moments''. Alternatively, a parametric distribution links different components and cells through a shared parameter set, allowing its parameters to be fitted using the same spectral basis. We call this treatment ``constrained moments''.

We demonstrate this statistical framework through accurate forward calculations and fits to diverse synchrotron SEDs. We compare the intensity fits with those obtained using empirical spectral laws, and demonstrate improved accuracy at low noise for spectra that depart from a curved power law, such as aged electron spectra \cite{kardashev1962nonstationarity, pacholczyk1970radio, jaffe1973dynamical}. We also show that retaining correlations through joint moments avoids biases that arise when assuming statistical independence.

We also test extrapolation from lower-frequency fits up to $300\,\mathrm{GHz}$ and find that the in-band spectral shape does not by itself fix the high-frequency departure from a power law. Reliable extrapolation requires either sufficiently broad frequency coverage and low noise in a free-moment fit, or a correctly specified electron-spectrum model (an ``electron family'') or empirical spectral law, whereas a mismatched family or spectral law can fit the band within the noise yet fail at higher frequencies.

The rest of the paper is organised as follows: Section~\ref{sec: single particle} derives the single-electron emission and propagation. Section~\ref{sec: geometry} forms the ensemble average and the channel response, and recovers the classical continuum limit. Section~\ref{sec: moment expansion} presents the spectral basis, the lattice of moments and linear fitting. Section~\ref{sec: classical models} presents the tests and their physical interpretation, and Sec.~\ref{sec: conclusions} discusses applicability.

\section{Single-particle emission and propagation}
\label{sec: single particle}

In this section, we construct the single-electron Stokes kernels (the emitted Stokes spectra per unit solid angle) used in the ensemble averages of Sec.~\ref{sec: geometry}. We first derive the harmonic emission from a helical orbit, express its polarisation in a fixed sky basis, and include Faraday rotation between the emitting element and the observer.

\begin{figure}[htbp]
 \centering
 \includegraphics[width=\textwidth]{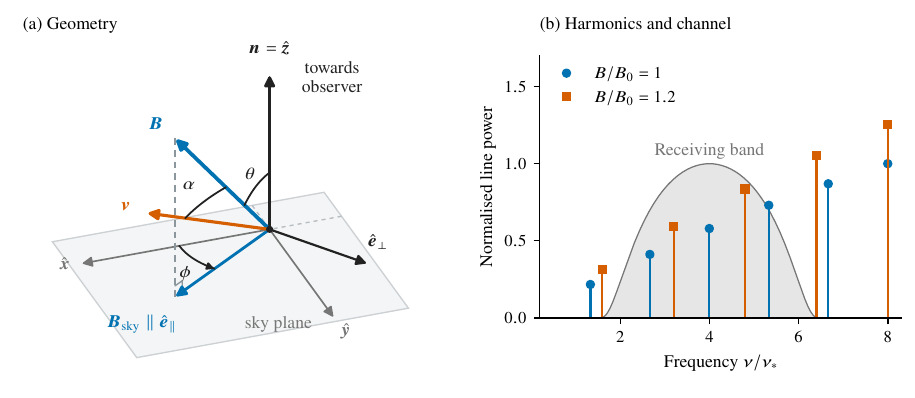}
 \\
 \caption{Emission geometry and harmonic lines in a fixed channel. (a)~Pitch angle $\alpha$, field inclination $\theta$ and projected-field azimuth $\phi$. (b)~Raising the field by $20\%$ raises the harmonic powers and moves the lines relative to the channel response, for $\gamma=20$ and $\mu=\eta=0.5$, with $\nu_\ast=eB_0/(2\pi\gamma m_ec)$.}
 \label{fig: angle geometry}
 \label{fig: moving lines}
 \vspace{-2mm}
\end{figure}

We use Gaussian cgs units, electron charge $q=-e$ with $e>0$, mass $m_e$, and $\boldsymbol\beta=\boldsymbol v/c$. For the fixed ray used in the derivation, the sky basis is right-handed: $\hat{\boldsymbol z}=\boldsymbol n$ points towards the observer, $\hat{\boldsymbol x}$ sets the position-angle reference, and $\hat{\boldsymbol y}=\boldsymbol n\times\hat{\boldsymbol x}$. The field makes an angle $\theta$ with $\boldsymbol n$, and its sky projection has azimuth $\phi$ measured from $\hat{\boldsymbol x}$ towards $\hat{\boldsymbol y}$. The electron pitch angle is $\alpha=\angle(\boldsymbol v,\boldsymbol B)$. We use $\mu=\cos\alpha$ and $\eta=\cos\theta$ below. Figure~\ref{fig: angle geometry}(a) shows these three angles and the two sky bases in a common spatial construction. Figure~\ref{fig: framework overview} lists all six parameters of the model.

\FloatBarrier

\subsection{Harmonic radiation from a helical orbit}
\label{sec: schott}

Here we derive the radiation\footnote{We use vacuum emission kernels. For a cold plasma with electron density $n_e$, this requires $\nu\gg\gamma\nu_p$, with electron plasma frequency $\nu_p\simeq9\,\mathrm{kHz}\,(n_e/\mathrm{cm}^{-3})^{1/2}$. Near the characteristic synchrotron frequency this gives $\nu\gg\nu_R$, with $\nu_R\simeq20\,\mathrm{Hz}\,(n_e/\mathrm{cm}^{-3})/[(B/\mathrm{G})\sin\theta]$ (see Ref.~\cite{ginzburg1965cosmic} and Sec.~8.3 of Ref.~\cite{rybicki1979radiative}).} from an electron on a prescribed helical orbit in a magnetic field uniform over the orbit, neglecting radiation backreaction~\cite{schott1912electromagnetic,trubnikov1958magnetic,ginzburg1965cosmic,rybicki1979radiative}. The field must be uniform and cooling negligible over the radiation-forming segment~\cite{2026PhyU...69..406B,ginzburg1965cosmic,pacholczyk1970radio}. Differences in magnetic field between emitting elements enter through the ensemble average. Long-term cooling enters only through the supplied electron distribution.

Both the emission and reception are described in a common inertial frame in which the macroscopic source and the observer are at rest. 
We denote emission and reception coordinate times by $t'$ and $t$.
For a constant field, $\gamma=(1-\beta^2)^{-1/2}$, $\beta_\parallel=\beta\cos\alpha$ and $\beta_\perp=\beta\sin\alpha$ are constant. The source-time gyrofrequency is $\omega_B=eB/(\gamma m_ec)$, with period $T_B=2\pi/\omega_B$. 
The reception-time interval is related to the emission-time interval by $\diff t/\diff t'=1-\boldsymbol n\cdot\boldsymbol\beta(t')$.\footnote{The factor comes from the change in light-travel time between emissions. Both intervals use coordinate time in the same inertial frame. A time-dilation factor would enter only if the emission interval were expressed in electron proper time.} Its average over one source-time period is $D=1-\beta\cos\alpha\cos\theta>0$, so one source-time orbit arrives over an interval $DT_B$. The observed harmonic frequencies are therefore~\cite{trubnikov1958magnetic,mahadevan1996harmony,leung2011numerical}
\begin{equation}
 \omega_m=\frac{m\omega_B}{D},\qquad
 m=1,2,\ldots.
 \label{eq: harmonic frequencies}
\end{equation}
The arrival-time relation and the Fourier normalisation of the far-zone radiation integral are standard~\cite{jackson1999classical,rybicki1979radiative}. The natural sky basis follows the projected field:
$\hat{\boldsymbol e}_\parallel=\cos\phi\,\hat{\boldsymbol x}+\sin\phi\,\hat{\boldsymbol y}$ and
$\hat{\boldsymbol e}_\perp=\boldsymbol n\times\hat{\boldsymbol e}_\parallel$.

\begin{samepage}
For the observer-time Fourier convention $e^{-i\omega_m t}$, the dimensionless polarisation amplitudes are
\begin{equation}
 \mathcal A^\parallel_m=(-1)^m\frac{\cos\theta-\beta_\parallel}{\sin\theta}J_m(z_m),
 \qquad
 \mathcal A^\perp_m=(-1)^m i\beta_\perp J_m'(z_m),\qquad
 z_m=\frac{m\beta_\perp\sin\theta}{D}.
 \label{eq: harmonic amplitudes}
\end{equation}
Here $J_m$ is the Bessel function of the first kind and a prime denotes differentiation with respect to the argument. The squared moduli of these amplitudes give the $I$ and $Q$ harmonic kernels of Refs.~\cite{leung2011numerical,pandya2016symphony}, and for $\beta_\parallel=0$ they reduce to the circular-orbit amplitudes~\cite{schott1912electromagnetic,schwinger1949classical,jackson1999classical}.
\end{samepage}

The powers per unit solid angle in harmonic $m$ are normalised to source time by dividing the energy emitted into that harmonic over one orbit by $T_B$:
\begin{equation}
    \frac{\diff P_m^\parallel}{\diff\Omega}=C_m^2|\mathcal A^\parallel_m|^2,\quad
    \frac{\diff P_m^\perp}{\diff\Omega}=C_m^2|\mathcal A^\perp_m|^2,
    \qquad
    C_m^2=\frac{e^2\omega_m^2}{2\pi cD}=\frac{e^2m^2\omega_B^2}{2\pi cD^3}.
    \label{eq: harmonic prefactor}
\end{equation}
The $D^{-3}$ normalisation follows from $\omega_m^2$ and the line Jacobian $\delta(\omega D-m\omega_B)=D^{-1}\delta(\omega-\omega_m)$ in the long-source-time Fourier limit.
Numerical normalisation checks reproduce the isotropic pitch-angle average of the Li\'enard power after summing over harmonics and integrating over all directions.

\subsection{Stokes parameters}
\label{sec: stokes harmonic}

We express the harmonic Stokes parameters in the natural basis $(\hat{\boldsymbol e}_\parallel,\hat{\boldsymbol e}_\perp)$ of Sec.~\ref{sec: schott}. The amplitudes in Eq.~\eqref{eq: harmonic amplitudes}, with the source-time normalisation $C_m^2$ of Eq.~\eqref{eq: harmonic prefactor}, give the Stokes powers in harmonic $m$~\cite{leung2011numerical,pandya2016symphony}\footnote{The $I$, $Q$ and $U$ expressions agree with Eqs.~(4)--(6) of \cite{pandya2016symphony}. The sign of $V$ follows the authors' 2017 correction to the accompanying code, which enforces their stated IEEE/IAU convention. }
\begin{subequations}\label{eq: harmonic stokes}
\begin{equation}
 I_m=\frac{\diff P_m^\parallel}{\diff\Omega}+\frac{\diff P_m^\perp}{\diff\Omega},\qquad
 Q_m=\frac{\diff P_m^\parallel}{\diff\Omega}-\frac{\diff P_m^\perp}{\diff\Omega},\qquad
 U_m=2C_m^2\mathrm{Re}(\mathcal A^\parallel_m\mathcal A^{\perp\ast}_m)=0,
 \label{eq: In}
\end{equation}
\begin{equation}
    V_m=-2C_m^2\mathrm{Im}(\mathcal A^\parallel_m\mathcal A^{\perp\ast}_m)
    =2C_m^2\frac{\beta_\perp(\cos\theta-\beta_\parallel)}{\sin\theta}J_m(z_m)J_m'(z_m).
    \label{eq: Vn}
\end{equation}
\end{subequations}
With this Fourier convention, positive $V$ corresponds to an electric vector rotating from $\hat{\boldsymbol e}_\parallel$ towards $\hat{\boldsymbol e}_\perp$ as observer time increases. Each nonzero single-particle harmonic is fully polarised~\cite{rybicki1979radiative}:
\begin{equation}
    I_m^2=Q_m^2+V_m^2.
    \label{eq: full pol}
\end{equation}
%
At $\theta=0,\pi$ the azimuth $\phi$, and thus the natural basis, is undefined, but the Stokes powers have finite limits: only $m=1$ radiates, with $I_1=C_1^2\beta_\perp^2/2$ and purely circular polarisation of opposite handedness at the two poles.

For $X\in\{I,Q,U,V\}$, let $X_m$ denote the single-electron source-time power per unit solid angle in harmonic $m$. At fixed $(\gamma,B,\mu,\eta)$, its \emph{spectral kernel} in the natural polarisation basis resolves this power in observer frequency $\nu$:
\begin{equation}
    \mathcal K^{\rm nat}_X(\nu)
    =\sum_{m=1}^{\infty}X_m\,\delta(\nu-\nu_m),
    \qquad \nu_m=\frac{\omega_m}{2\pi},
    \label{eq: single spectral kernel}
\end{equation}
which is a spectral density per unit solid angle. 

An observing channel integrates these lines over a fixed frequency band.
Each line contributes to a channel with a weight set by the channel response at the line frequency (Sec.~\ref{sec: channel response}).
Figure~\ref{fig: moving lines}(b) shows how varying the field strength changes the line powers, frequencies and channel weights.

\subsection{Faraday rotation of the emitted polarisation}
\label{sec: single element propagation}


Faraday rotation acts on the radiation propagating from each emitting element to the observer.
We write the subscript $P$ for the combination $Q+iU$ of any Stokes quantity, with
$\mathcal K_P^{\rm nat}\equiv\mathcal K_Q^{\rm nat}$ because $U_m=0$ in the natural basis, and
use the spin weights $\mathsf s_I=\mathsf s_V=0$ and $\mathsf s_P=2$.
Along a straight ray, let $s$ increase towards the observer, $n_e$ be the thermal-background free-electron density and $B_\parallel$ the field component positive towards the observer. The high-frequency cold-plasma expression for the Faraday depth from the element to the observer is~\cite{rybicki1979radiative,burn1966depolarization}
\begin{equation}
    \Psi\simeq0.812\int_{\rm emitter}^{\rm observer}
    \left(\frac{n_e(s)}{\mathrm{cm^{-3}}}\right)
    \left(\frac{B_\parallel(s)}{\mu\mathrm G}\right)
    \diff\!\left(\frac{s}{\mathrm{pc}}\right)\,\mathrm{rad\,m^{-2}}.
    \label{eq: RM}
\end{equation}
We retain pure Faraday rotation, neglecting absorption, scattering, refraction and conversion~\cite{pandya2016symphony}. At vacuum wavelength $\lambda=c/\nu$, expressed in metres in Faraday phases, the position angle changes by $\lambda^2\Psi$. Rotating from the natural basis into the fixed
observer basis $(\hat{\boldsymbol x},\hat{\boldsymbol y})$ through the projected-field azimuth
$\phi$ (Fig.~\ref{fig: angle geometry}(a))~\citep{Chandrasekhar1960book, rybicki1979radiative}
adds the angle $\phi$, so an element at Faraday depth $\Psi$ contributes
\begin{equation}
    e^{i\mathsf s_X(\phi+\lambda^2\Psi)}\,\mathcal K_X^{\rm nat}(\nu),
    \qquad X\in\{I,P,V\},
    \label{eq: kernel hierarchy}
\end{equation}
to the observed sky-basis spectrum.

We take $\Psi$ to be frequency independent and exclude ion and relativistic-plasma corrections and Faraday conversion~\cite{huang2011faraday}.  Section~\ref{sec: screen model} averages these propagated contributions over emitting sites.

\section{From electron ensembles to observed emission}
\label{sec: geometry}

In this section, we average the single-electron kernels of Sec.~\ref{sec: single particle} over the joint distribution of emitting parameters and Faraday depth, and integrate over the observing channels. The discussion in this section provides the starting point for the spectral basis in Sec.~\ref{sec: moment expansion}. We also recover the classical results for intensity and linear polarisation in the ultra-relativistic continuum limit.

Throughout the formalism, all quantities are understood to refer to a fixed ray propagating towards the observer along direction $\boldsymbol n$, with distance coordinate $s$ defined as in Sec.~\ref{sec: single element propagation}. We therefore suppress the explicit dependence on $\boldsymbol n$.
A density $\mathcal P_\star(\boldsymbol u\mid\boldsymbol q)$ is normalised over $\boldsymbol u$ at fixed $\boldsymbol q$, and its subscript names the distribution (Table~\ref{tab: averages}).
The average of $g$ over it is $\langle g\rangle_\star=\int g\,\mathcal P_\star\diff\boldsymbol u$, which includes cases when $g$ depends on only some of $\boldsymbol u$.

\begin{table}[tbp]
\centering
\small
\caption{Probability densities used in the ensemble averages.}
\label{tab: averages}
\smallskip
\begin{tabular}{@{}llll@{}}
\toprule
Density & Normalised over & Held fixed & Reference \\
\midrule
$\mathcal P_\mu$ & $\mu$ & $\gamma,B,\eta,\phi$ & Eq.~\eqref{eq: beaming limit} \\
$\mathcal P_{\rm col}$ & $\boldsymbol p,\phi$ & --- & Eq.~\eqref{eq: joint emission depth} \\
$\mathcal P_{\rm F}$ & $\boldsymbol p,\phi,\Psi$ & --- & Eq.~\eqref{eq: joint emission depth} \\
$\mathcal P_{\rm scr}$ & $\Psi$ & --- & Sec.~\ref{sec: independent screen limit} \\
\bottomrule
\end{tabular}
\end{table}

\subsection{Source ensemble and intrinsic emission}
\label{sec: local population}

We assume a statistically stationary electron distribution over the emission times sampled by the sightline and observing interval. The distribution is independent of electron gyrophase, and emission from different electrons is incoherent, so Stokes powers add~\cite{pandya2016symphony}. 
Coherent emission, source variability over these emission times and spatial weighting that selects orbital phases within an emitting segment lie outside this reference model, and their discrepancies are not bounded here.

We group the four intrinsic emission parameters in $\boldsymbol p=(\gamma,B,\mu,\eta)$ and keep $\phi$ separate, since it affects only the polarisation angle (Eq.~\eqref{eq: kernel hierarchy}). The parameter ranges are $\gamma\geq1$, $B\geq0$, $\mu,\eta\in[-1,1]$ and $\phi\in[0,2\pi)$. Let $s$ increase from the source boundary at $0$ towards the observer at $s=L_s$. The source-electron number density $n_{\rm src}(s,\boldsymbol p,\phi)$ is defined per unit volume and per unit $\diff\boldsymbol p\,\diff\phi$, with $\diff\boldsymbol p=\diff\gamma\,\diff B\,\diff\mu\,\diff\eta$. The electrons are counted in fixed volume elements on a slice of constant emission time $t'$.
Magnetic field magnitudes are weighted by the number of emitting electrons at each value. Energy, pitch angle and field properties can be correlated. 

Integrating the single-electron contributions $e^{i\mathsf s_X\phi}\,\mathcal K_X^{\rm nat}(\nu)$ against the number density $n_{\rm src}$ gives the local emissivity $j_X(\nu,s)$.\footnote{Because $\mathcal K_X^{\rm nat}(\nu)$ gives power per unit solid angle in the observer's direction, no $1/(4\pi)$ factor is needed. A volume element at distance $R$ from the observer subtending $\diff\Omega$ has volume $R^2\diff\Omega\diff s$. The inverse-square factor in the received flux cancels $R^2$, and dividing by $\diff\Omega$ then gives the intensity contribution $j_X(\nu)\diff s$.}
\begin{equation}
    j_X(\nu,s)=\int n_{\rm src}(s,\boldsymbol p,\phi)\,e^{i\mathsf s_X\phi}\mathcal K_X^{\rm nat}(\nu,\boldsymbol p)\diff\boldsymbol p\,\diff\phi.
    \label{eq: los intensity}
\end{equation}
For a stationary population, the same integral gives the energy directed towards the observer per unit reception time, frequency, solid angle and source volume, since the light-travel time from a fixed volume element is constant~\cite{scheuer1968synchrotron} (see also Sec.~6.7 of Ref.~\cite{rybicki1979radiative}).
Existing emissivity calculations treat general energy and pitch-angle distributions~\cite{petrosian1981synchrotron}, thermal cyclotron--synchrotron spectra~\cite{mahadevan1996harmony,leung2011numerical}, and thermal, power-law and $\kappa$ distributions~\cite{pandya2016symphony}.

Equation~\eqref{eq: los intensity} gives the emissivity of a medium at rest. For a medium in steady bulk motion, the Doppler shift acts on the kernels as a rescaling of the field strength and amplitude, and aberration changes the field angles seen along the ray.
This could be treated by applying the boost operator associated with this transformation to ensembles of directions~\cite{chluba2026boost,hoey2026derivation}, but is not included here.

\subsection{Faraday rotation with distributed emission}
\label{sec: screen model}

The polarisation of each emitting element rotates according to the Faraday depth \cite{burn1966depolarization} between its emission site and the observer.
Pure rotation depends only on the Faraday depth of each emission site, so the order of the sites along the ray does not enter.
Absorption and Faraday conversion depend on that order and require a general transfer calculation~\cite{pandya2016symphony,huang2011faraday}, which we leave to future work. 

For a specified plasma realisation, let $\Psi(s)$ be the Faraday depth of site $s$, defined by Eq.~\eqref{eq: RM}. The joint probability density of emitting parameters and Faraday depth, per unit $\diff\boldsymbol p\,\diff\phi\,\diff\Psi$, is
\begin{equation}
    \mathcal P_{\rm F}(\boldsymbol p,\phi,\Psi)
    =\frac{1}{\mathcal N_{\rm src}}\int_0^{L_s}
    n_{\rm src}(s,\boldsymbol p,\phi)\delta\!\left[\Psi-\Psi(s)\right]\diff s,
    \qquad
    \int\mathcal P_{\rm F}(\boldsymbol p,\phi,\Psi)\diff\Psi
    =\mathcal P_{\rm col}(\boldsymbol p,\phi),
    \label{eq: joint emission depth}
\end{equation}
where $\mathcal N_{\rm src}=\int_0^{L_s}\!\int n_{\rm src}\,\diff\boldsymbol p\,\diff\phi\,\diff s>0$ is the source-electron column. The second equation defines the column probability density $\mathcal P_{\rm col}$ per unit $\diff\boldsymbol p\,\diff\phi$.
Applying Eq.~\eqref{eq: kernel hierarchy} to each emitting site gives the source contribution\footnote{We assume that no radiation enters the ray at its far end. Incident radiation would add a separate term, rotated by the Faraday depth of the whole ray.}
\begin{equation}
    X_\nu^{\rm obs}
    =\int_0^{L_s} j_X(\nu,s)\,e^{i\mathsf s_X\lambda^2\Psi(s)}\diff s
    =\mathcal N_{\rm src}
    \left\langle e^{i\mathsf s_X(\phi+\lambda^2\Psi)}\mathcal K_X^{\rm nat}(\nu,\boldsymbol p)\right\rangle_{\rm F}.
    \label{eq: distributed emission average}
\end{equation}
When intrinsic parameters, field orientation and Faraday depth are correlated, the sky and Faraday factors of Eq.~\eqref{eq: kernel hierarchy} do not factorise in this average. 
Contributions rotated by different angles add as vectors in the $Q$--$U$ plane and can reduce $|P^{\rm obs}|$ (Fig.~\ref{fig: depolarisation}(b)).

\subsection{Observer-channel response}
\label{sec: channel response}
The ideal spectrum of Eq.~\eqref{eq: single spectral kernel} consists of Dirac lines whose frequencies move with $\gamma$, $B$ and the angles (Fig.~\ref{fig: moving lines}(b)). A radio receiver integrates each Stokes spectrum over frequency against the channel response $R_j(\nu)$, taken here to be the same for all Stokes components, and Sec.~\ref{sec: moment expansion} expands these channel integrals rather than the spectra.
The harmonic powers of Sec.~\ref{sec: stokes harmonic} give the Faraday-rotated channel kernels
\begin{align}
    \mathcal K_{Xj}(\boldsymbol p,\Psi)
    &\equiv\int R_j(\nu)\mathcal K^{\rm nat}_X(\nu,\boldsymbol p)
    e^{i\mathsf s_X(c/\nu)^2\Psi}\diff\nu\nonumber\\
    &=\sum_{m\geq1}X_mR_j(\nu_m)e^{i\mathsf s_X(c/\nu_m)^2\Psi},
    \qquad X\in\{I,P,V\},
    \label{eq: channel kernel}
\end{align}
with $P_m\equiv Q_m$. The kernels for $I$ and $V$ are independent of $\Psi$. With Eq.~\eqref{eq: distributed emission average}, the channel values are
\begin{equation}
    X_j^{\rm obs}=\int R_j(\nu)X_\nu^{\rm obs}\diff\nu
    =\mathcal N_{\rm src}\langle e^{i\mathsf s_X\phi}\mathcal K_{Xj}\rangle_{\rm F}.
    \label{eq: full channel average}
\end{equation}
Each harmonic carries its own channel weight and its own Faraday phase. 
The channel-response normalisation determines whether the channel values are band integrated or band averaged.

Equations~\eqref{eq: channel kernel} and \eqref{eq: full channel average} hold for any channel response.
A parameter derivative of the channel kernel differentiates both the line powers and their weights $R_j(\nu_m)$, so the expansion presented in Sec.~\ref{sec: moment expansion} requires smooth channel responses.
The channel response must also vanish near zero frequency, where Faraday rotation grows without bound.
The true channel response can be approximated by a weighted sum of narrower channel responses that meet these conditions. Because the channel kernel is linear in $R_j$, the same weights combine the corresponding kernels.

\begin{figure}[tbp]
\centering
\includegraphics[width=\textwidth]{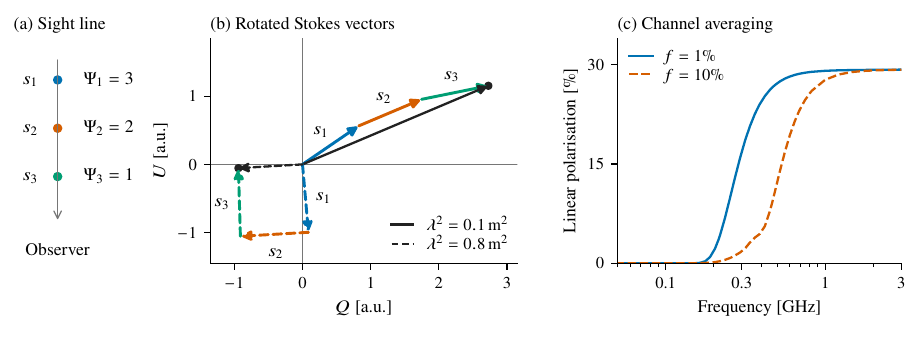}
\caption{Faraday depolarisation. (a)~Three emission sites at different Faraday depths. (b)~Their polarisation vectors, rotated by $2\lambda^2\Psi_i$, add head to tail and partly cancel at the larger $\lambda^2$. (c)~Wider channels average over more rotation and depolarise up to higher frequencies.}
\label{fig: depolarisation}
\end{figure}

The examples in this paper use one such channel response, a smooth unit-area bump centred on $\nu_j$ with fractional rms width $f$ (Eq.~\eqref{eq: forward response families}). It is nonzero only within $\nu_j(1\pm2.5f)$, so it vanishes near zero frequency for $f<0.4$.
Faraday rotation changes the polarisation angle across a channel, and averaging these differently oriented contributions reduces the measured polarisation (Fig.~\ref{fig: depolarisation}(c)). Widening the channels from $f=1\%$ to $f=10\%$ raises the frequency at which the polarisation falls to half of its high-frequency value from $0.28$ to $0.53\,\mathrm{GHz}$. The ensemble in this example has $\gamma^{-2.6}$ electrons on $[300,10^5]$, fields uniform on $[2,8]\,\mu\mathrm G$, isotropic pitch angles, $\eta$ uniform on $[-1/2,1/2]$, $\langle e^{2i\phi}\rangle_{\rm F}=0.4$ and Gaussian Faraday depths of mean $8$ and standard deviation $0.5\,\mathrm{rad\,m^{-2}}$, all independent. 

\begin{figure}[!t]
\centering
\includegraphics[width=\textwidth]{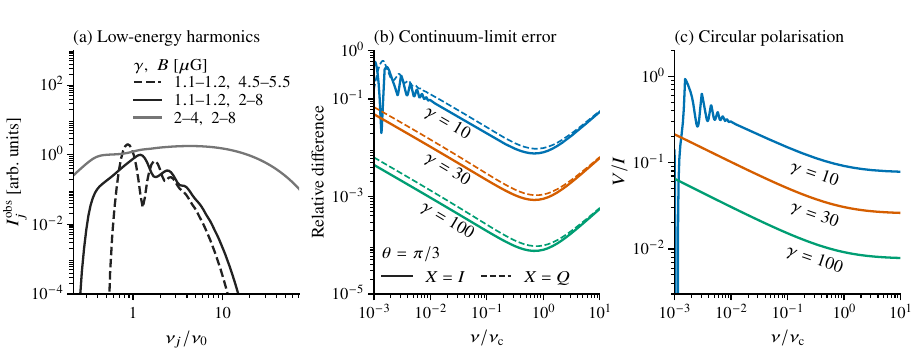}
\caption{Harmonic structure and the continuum limit. (a)~At low energy, individual harmonics remain visible in the channel intensities, with $\nu_0=eB_0/(2\pi m_ec)$ and $B_0=5\,\mu$G. (b)~The relative error of the continuum kernels in $I$ and $Q$ falls as $\gamma^{-2}$ at fixed $\nu/\nu_{\rm c}$, where $\nu_{\rm c}=3eB_\perp\gamma^2/(4\pi m_ec)$ is the characteristic frequency. At fixed $\gamma$ it is smallest just below $\nu_{\rm c}$. (c)~The circular polarisation $V/I$ omitted by the continuum kernels.
Each harmonic spans a range of frequencies because its frequency depends on pitch angle. At small $\gamma\nu$, only a few harmonics contribute at any given frequency. Sharp features appear as harmonics start or stop contributing. These curves are evaluated without channel averaging. $V$ changes sign near $\nu=10^{-3}\nu_{\rm c}$.}
\label{fig: harmonic continuum}
\end{figure}

The channel spectra above describe one fixed line of sight with a common scalar channel response for all Stokes components. A telescope also averages over sky directions with its beam, mixes the Stokes components, rotates the polarisation into the reported sky basis and applies a calibration. All of these operations are linear in the Stokes spectra. Averaging over sky directions changes the spectral shape~\cite{chluba2017rethinking,remazeilles2021peeling} and is not included in the basis or the examples of this paper.

\subsection{Ultra-relativistic continuum limit}
\label{sec: continuum limit}

For $\gamma\gg1$, the pitch-angle average of the harmonic kernels, taken over frequency intervals containing many harmonics, reduces to the classical synchrotron continuum for intensity and linear polarisation~\cite{westfold1959polarization,ginzburg1965cosmic,legg1968elliptical,rybicki1979radiative}. Our formalism does not depend on this limit. We use it to check the harmonic kernels against the classical results (Fig.~\ref{fig: harmonic continuum}) and, in Sec.~\ref{sec: classical models}, as the kernel for electrons with $\gamma\geq300$.

\FloatBarrier

At fixed $\nu/\gamma^2$, relativistic beaming confines the emission towards the observer to electrons with $|\alpha-\theta|\lesssim1/\gamma$.\footnote{The continuum kernels assume $\gamma\sin\theta\gg1$. Lines of sight within about $1/\gamma$ of the field, or pitch-angle densities concentrated at $\alpha\lesssim1/\gamma$, need the small-pitch-angle treatment of Ref.~\cite{epstein1973synchrotron}.}
Let $\mathcal P_\mu(\mu)$ denote the conditional pitch-angle density $\mathcal P_\mu(\mu\mid\gamma,B,\eta,\phi)$ of the column ensemble $\mathcal P_{\rm col}$ in Eq.~\eqref{eq: joint emission depth}. If the density varies little within the narrow beaming cone, the $\mu$-integral evaluates it at $\mu=\eta$:
\begin{equation}
\begin{split}
    \int_{-1}^{1}\mathcal P_\mu(\mu)\,\mathcal K_X^{\rm nat}(\nu,\gamma,B,\mu,\eta)\diff\mu
    & \simeq
    \mathcal P_\mu(\eta) \int_{-1}^{1}\mathcal K_X^{\rm nat}(\nu,\gamma,B,\mu,\eta)\diff\mu \\
    & \equiv 2\mathcal P_\mu(\eta)\,\mathcal K_X^{\rm cont}(\nu,\gamma,B,\eta),
    \qquad X\in\{I,Q\},
\end{split} \label{eq: beaming limit}
\end{equation}
which defines the continuum kernel $\mathcal K_X^{\rm cont}=\tfrac12\int_{-1}^{1}\mathcal K_X^{\rm nat}\diff\mu$, the pitch-angle average of the harmonic kernel. In the natural basis, $U_m=0$ in Eq.~\eqref{eq: In} gives $\mathcal K_U^{\rm cont}=0$. The pitch integral of $V$ vanishes at leading order in $1/\gamma$. We thus neglect $\mathcal K_V^{\rm cont}$. Appendix~\ref{sec: continuum derivation} evaluates the pitch integral from the large-order forms of $J_m$ and $J_m'$ with the source-time normalisation of Eq.~\eqref{eq: harmonic prefactor}. With $B_\perp=B\sin\theta$, the characteristic frequency $\nu_{\rm c}=3eB_\perp\gamma^2/(4\pi m_ec)$ and the fixed $(\gamma,B,\eta)$ arguments suppressed, its leading order in $1/\gamma$ at fixed $x$ gives the classical kernels \citep{Blumenthal1970, rybicki1979radiative}
\begin{equation}
    \begin{aligned}
    \mathcal K_I^{\rm cont}(\nu)&\simeq A_\theta F(x),&
    \mathcal K_Q^{\rm cont}(\nu)&\simeq-A_\theta G(x),&
    \qquad A_\theta&=\frac{\sqrt3e^3B_\perp}{4\pi m_ec^2},\\
    F(x)&=x\int_x^\infty K_{5/3}(\xi)\diff\xi,&
    G(x)&=xK_{2/3}(x),&
    x&=\frac{\nu}{\nu_{\rm c}}.
    \end{aligned}
    \label{eq: directional continuum}
\end{equation}
Here $K_{5/3}$ and $K_{2/3}$ are modified Bessel functions of the second kind, and the sign of $Q$ follows Sec.~\ref{sec: stokes harmonic}. The continuum kernels used in the rest of the paper are these leading-order forms. 

In Fig.~\ref{fig: harmonic continuum}, individual harmonics remain resolved in the channel spectra at low energy (panel~a), and panel~(b) gives the error of the continuum kernels of Eq.~\eqref{eq: directional continuum} in $I$ and $Q$, their relative difference from the pitch-angle average of the harmonic kernels.
At fixed $\gamma$, the error falls as $\nu^{-2/3}$ below $\nu_{\rm c}$ and is smallest, at about $\gamma^{-2}$, just below $\nu_{\rm c}$. Well below $\nu_{\rm c}$ it depends on $\gamma$ and $\nu$ only through the order of the harmonics emitting at $\nu$, which is proportional to $\gamma\nu$. Above $\nu_{\rm c}$ the error grows only in proportion to $\nu/(\gamma^2\nu_{\rm c})$, while the emission itself falls exponentially.
Panel~(c) shows the $V$ omitted by the continuum kernels.

\paragraph{Higher-order corrections in $1/\gamma$.}
The continuum kernels of Eq.~\eqref{eq: directional continuum} retain only the leading order in $1/\gamma$. The omitted circular polarisation $V$ enters at first order in $1/\gamma$~\cite{legg1968elliptical}, and vanishes for the test populations of Sec.~\ref{sec: classical models} because their inclination and pitch-angle distributions are even (Eq.~\eqref{eq: angular parity}). The corrections to $I$ and $Q$ are of second order in $1/\gamma$. Weighting the $\gamma^{-2}$ errors of Fig.~\ref{fig: harmonic continuum}(b) by the electron and field distributions of the test populations with isotropic pitch angles, we estimate that they change $I$ and $Q$ by less than $10^{-6}$ of the intensity over $0.1$--$300\,\mathrm{GHz}$.

\paragraph{Connections to classical formulae.}
The intensity kernel $\mathcal K_I^{\rm cont}$ of Eq.~\eqref{eq: directional continuum} equals the angle-integrated single-electron spectrum of Rybicki \& Lightman~\cite{rybicki1979radiative} divided by $4\pi$ (their Eq.~6.33 at $\alpha=\theta$, converted with $P_\nu=2\pi P_\omega$). The ratio $-\mathcal K_Q^{\rm cont}/\mathcal K_I^{\rm cont}=G(x)/F(x)$ is the classical linear-polarisation fraction for electrons of a single energy.
Integrating Eq.~\eqref{eq: directional continuum} over a power-law electron density $\diff\mathcal N/\diff\gamma\propto\gamma^{-p}$ that extends well beyond the emitting energies gives the intrinsic spectrum $I_\nu\propto B_\perp^{(p+1)/2}\nu^{-(p-1)/2}$ and linear-polarisation fraction $|Q_\nu|/I_\nu=(p+1)/(p+7/3)$ in a uniform field~\cite{westfold1959polarization,ginzburg1965cosmic,rybicki1979radiative}.
With isotropic pitch angles the same kernels give the regular-field emissivity, and their average over isotropic field directions gives the random-field kernel~\cite{crusius1986synchrotron,ghisellini1988synchrotron} used in \textsc{GALPROP}~\cite{strong2011electron,orlando2013synchrotron}.


\section{A spectral basis for synchrotron emission}
\label{sec: moment expansion}

Section~\ref{sec: geometry} gives the Stokes emission of a single electron after Faraday rotation and channel integration as a function of $(\gamma,B,\mu,\eta,\phi,\Psi)$. The observed channel spectrum follows by integrating this emission over the joint population distribution and multiplying by the source-electron column (Eq.~\eqref{eq: full channel average}). In this section, we approximate this integral by a finite linear combination of basis spectra, with coefficients set by ``generalised moments'' shared by all channels (Fig.~\ref{fig: framework overview}).

We build the spectral basis from the kernels and their responses to changes in the parameters. Parameters that leave the spectral shape unchanged can be integrated out, with any overall factor they introduce included in the moment weights. The two kernels lead to different choices of variables.
\begin{itemize}
    \item \textit{Harmonic kernels (Sec.~\ref{sec: joint general}).}
    The harmonic calculation resolves the emission in individual lines, which can be necessary at low frequencies. The normalised intrinsic spectra depend on $(\gamma,B,\mu,\eta)$. Three combinations of these parameters set the line spacing and relative Stokes line strengths in the natural basis (Eq.~\eqref{eq: harmonic shape combinations}). Although a basis in these combinations is more compact, a basis in the physical parameters is more explicit and makes model conditions easier to impose. For example, assuming isotropic pitch angles simply fixes the $\mu$ distribution. \textsc{SyncEM} therefore supports both representations, and Sec.~\ref{sec: joint general} presents the default construction in the physical parameters.
    \item \textit{Continuum kernels (Sec.~\ref{sec: joint relativistic}).}
    When the conditions of Sec.~\ref{sec: continuum limit} hold, the continuum kernels give a faster calculation of intensity and linear polarisation. Relativistic beaming localises the pitch-angle average near $\mu=\eta$, and the average of $V$ vanishes at the order retained. The intrinsic spectral shape depends on $(\gamma,B,\eta)$, but only through the characteristic frequency $\nu_{\rm c}\propto B_\perp\gamma^2$. The basis can again be built in the physical parameters or, more compactly, with $\ln\nu_{\rm c}$ in place of $(\gamma,B,\eta)$, and \textsc{SyncEM} supports both. Section~\ref{sec: joint relativistic} presents the compact construction.
\end{itemize}

In both cases, Faraday depth $\Psi$ produces a frequency-dependent rotation and enters the basis for linear polarisation. The azimuth $\phi$ acts on $P=Q+iU$ through the frequency-independent factor $e^{2i\phi}$. It therefore enters only the moment coefficients, and the basis spectra are independent of $\phi$.

With a kernel chosen and a spectral basis constructed for the population, or for its low-energy (harmonic) and high-energy (continuum) parts when both kernels are used, Sec.~\ref{sec: coarse graining} divides the remaining parameter space into a grid of cells and combines their local expansions. It also shows how specified population distributions constrain the moments.

\begin{samepage}
Integrating the basis spectra over an experiment's channel responses reduces each spectrum to one value per channel and Stokes component. Once we build this basis at fixed reference points, we can rapidly predict the Stokes spectra that the experiment would observe for any synchrotron population model covered by the expansion. Each prediction uses the model's moments and a single matrix--vector product. To make simulations and SED fitting more efficient, we can further compress the linear representation from the moment expansion. Methods such as Gram--Schmidt orthogonalisation and principal component analysis (PCA) can reorganise the channel basis. For linear SED fitting, Sec.~\ref{sec: linear fit} uses a singular-value decomposition (SVD) of the noise-weighted basis to select the moment combinations to retain.
\par
\end{samepage}

\subsection{Spectral basis via harmonic kernels}
\label{sec: joint general}

We first expand the harmonic channel kernel in Legendre polynomials of the angular cosines $\mu$ and $\eta$, then Taylor-expand the projected kernels in $\gamma$, $B$ and $\Psi$. A local Taylor expansion in the angular cosines can be inefficient when their distributions are smooth and broad across $[-1,1]$ and the kernel varies rapidly with angle. Legendre polynomials can represent this dependence compactly over the full interval.

The angular expansion and its coefficients are
\begin{equation}
    \begin{aligned}
    \mathcal K_{Xj}(\boldsymbol p,\Psi)
    &=\sum_{\ell,k\geq0}\mathcal K_{Xj\ell k}(\gamma,B,\Psi)\,\Leg_\ell(\mu)\Leg_k(\eta),\\
    \mathcal K_{Xj\ell k}(\gamma,B,\Psi)
    &=\frac{(2\ell+1)(2k+1)}4\int_{-1}^{1}\!\int_{-1}^{1}
    \mathcal K_{Xj}(\boldsymbol p,\Psi)
    \Leg_\ell(\mu)\Leg_k(\eta)\diff\mu\diff\eta.
    \end{aligned}
    \label{eq: legendre spectral response}
\end{equation}
Inserting the Legendre expansion of Eq.~\eqref{eq: legendre spectral response} into Eq.~\eqref{eq: full channel average} and integrating over $\mu$, $\eta$ and $\phi$ gives
\begin{equation}
    \begin{aligned}
    X_j^{\rm obs}
    &=\mathcal N_{\rm src}\sum_{\ell,k\geq0}\int
    \mathcal W^{(\mathsf s_X)}_{\ell k}\,\mathcal K_{Xj\ell k}\,
    \diff\gamma\,\diff B\,\diff\Psi,\\
    \mathcal W^{(\mathsf s_X)}_{\ell k}(\gamma,B,\Psi)
    &\equiv\int\mathcal P_{\rm F}(\boldsymbol p,\phi,\Psi)\,e^{i\mathsf s_X\phi}\,
    \Leg_\ell(\mu)\Leg_k(\eta)\,\diff\mu\,\diff\eta\,\diff\phi.
    \end{aligned}
    \label{eq: reduced general}
\end{equation}
The reduced distribution $\mathcal W^{(\mathsf s_X)}_{\ell k}$ equals the marginal density of $(\gamma,B,\Psi)$ multiplied by the conditional average of $e^{i\mathsf s_X\phi}\Leg_\ell(\mu)\Leg_k(\eta)$ at fixed $(\gamma,B,\Psi)$. For $\ell=k=0$ and $I$ it is the density itself.

We Taylor-expand the projected channel kernels $\mathcal K_{Xj\ell k}(\gamma,B,\Psi)$ of Eq.~\eqref{eq: legendre spectral response} in energy, field magnitude and Faraday depth. Put $\Delta\gamma=\gamma-\gamma_0$, $\Delta B=B-B_0$ and $\Delta\Psi=\Psi-\Psi_{\rm ref}$, with fixed reference values $\gamma_0>1$, $B_0>0$ and $\Psi_{\rm ref}$. The Taylor exponents $\boldsymbol r=(r_\gamma,r_B,r_\Psi)$ are nonnegative integers with total degree $|\boldsymbol r|=r_\gamma+r_B+r_\Psi$. We write $\Delta^{\boldsymbol r}=(\Delta\gamma)^{r_\gamma}(\Delta B)^{r_B}(\Delta\Psi)^{r_\Psi}$, $\partial^{\boldsymbol r}=\partial_\gamma^{r_\gamma}\partial_B^{r_B}\partial_\Psi^{r_\Psi}$ and $\boldsymbol r!=r_\gamma!\,r_B!\,r_\Psi!$. 
To total degree $N$, the spectral basis expansion and its basis functions are
\begin{equation}
    \mathcal K_{Xj\ell k}(\gamma,B,\Psi)
    \simeq\sum_{|\boldsymbol r|\leq N}\widetilde X_{j\ell k\boldsymbol r}\,\Delta^{\boldsymbol r},
    \qquad
    \widetilde X_{j\ell k\boldsymbol r}
    =\left.\frac{\partial^{\boldsymbol r}
    \mathcal K_{Xj\ell k}}{\boldsymbol r!}\right|_{\gamma_0,B_0,\Psi_{\rm ref}}.
    \label{eq: channel basis expansion}
\end{equation}
The spectral basis functions $\widetilde X_{j\ell k\boldsymbol r}$ depend on the radiation model, channel and reference values, not on the ensemble. For $X\in\{I,V\}$ the kernel is independent of $\Psi$, so $\widetilde X_{j\ell k\boldsymbol r}=0$ for $r_\Psi>0$.

The derivatives in Eq.~\eqref{eq: channel basis expansion} act on each harmonic term of Eq.~\eqref{eq: channel kernel}, including its channel weight $R_j(\nu_m)$, because the line frequency $\nu_m$ moves with $\gamma$ and $B$. They therefore require the smooth channel responses of Sec.~\ref{sec: channel response}. For a channel response that vanishes outside a finite frequency band, an upper limit on $\gamma$ and a positive lower limit on $B$ leave finitely many harmonics in the band (Eq.~\eqref{eq: harmonic frequencies}), and separation from $\gamma=1$ keeps the energy derivatives finite.

Truncating Eq.~\eqref{eq: reduced general} at degrees $L_\mu$ and $L_\eta$ and inserting the Taylor expansion of Eq.~\eqref{eq: channel basis expansion} gives the channel prediction and the moments $M^{(0)}$ and $M^{(2)}$:
\begin{equation}
    \begin{aligned}
    X_j^{\rm obs}
    &\simeq\mathcal N_{\rm src}
    \sum_{0\leq\ell\leq L_\mu} \sum_{\ 0\leq k\leq L_\eta}
    \sum_{|\boldsymbol r|\leq N}
    \widetilde X_{j\ell k\boldsymbol r}M^{(\mathsf s_X)}_{\ell k\boldsymbol r},\\
    M^{(\mathsf s_X)}_{\ell k\boldsymbol r}
    &\equiv\int\mathcal W^{(\mathsf s_X)}_{\ell k}\,\Delta^{\boldsymbol r}\,
    \diff\gamma\,\diff B\,\diff\Psi
    =\left\langle e^{i\mathsf s_X\phi}\Leg_\ell(\mu)\Leg_k(\eta)\,\Delta^{\boldsymbol r}\right\rangle_{\rm F}.
    \end{aligned}
    \label{eq: finite joint response}
\end{equation}
We call the $M^{(\mathsf s_X)}_{\ell k\boldsymbol r}$ joint generalised moments, which retain correlations between the parameters.

\begin{figure}[!t]
\centering
\includegraphics[width=\textwidth]{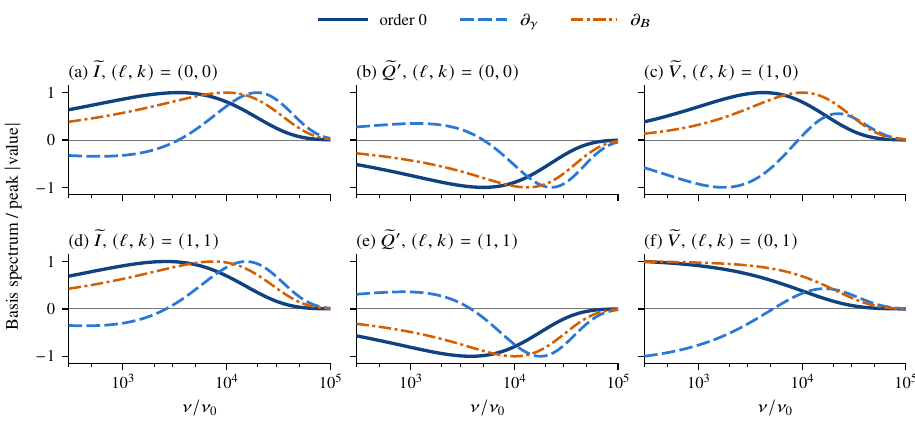}
\caption{Harmonic basis spectra of one cell at $\gamma_0=100$ and $B_0=5\,\mu$G without Faraday rotation, each divided by its peak, with $\nu_0=eB_0/(2\pi m_ec)$. $Q'$ is the linear polarisation in the natural basis. (a,\,b,\,d,\,e)~$I$ and $Q'$ match the continuum basis spectra. (c,\,f)~The finite beaming cone gives $V$ for odd $\ell+k$.}
\label{fig: harmonic basis spectra}
\end{figure}

To illustrate the spectral basis, Fig.~\ref{fig: harmonic basis spectra} shows the order-zero basis spectra and those for the first derivatives in $\gamma$ and $B$ at one reference point. Appendix~\ref{sec: continuum derivation} reports the angular-quadrature checks and compares the spectra with the continuum result.

\paragraph{Vanishing modes and Faraday derivatives.}
Some basis spectra vanish for every ensemble.
Under simultaneous reversal $(\mu,\eta)\mapsto(-\mu,-\eta)$, the harmonic line positions are unchanged, $I_m,Q_m$ are even and $V_m$ is odd. Hence
\begin{equation}
    \widetilde I_{j\ell k\boldsymbol r}=\widetilde P_{j\ell k\boldsymbol r}=0
    \quad(\ell+k\ \mathrm{odd}),\qquad
    \widetilde V_{j\ell k\boldsymbol r}=0\quad(\ell+k\ \mathrm{even}).
    \label{eq: angular parity}
\end{equation}
In Eq.~\eqref{eq: channel kernel}, each derivative with respect to Faraday depth multiplies the $m$th term of the linear-polarisation kernel by $2i(c/\nu_m)^2$.

\paragraph{Alternative expansions.}
The harmonic basis can also be constructed using three effective parameters that determine the intrinsic spectral shape. For $\beta_\perp\sin\theta\ne0$, Eqs.~\eqref{eq: harmonic frequencies} and \eqref{eq: harmonic amplitudes} give
\begin{equation}
    \nu_1=\frac{\omega_B}{2\pi D},\qquad
    \frac{z_m}{m}=\frac{\beta_\perp\sin\theta}{D},\qquad
    r_{\rm pol}\equiv\frac{\cos\theta-\beta_\parallel}{\beta_\perp\sin\theta}.
    \label{eq: harmonic shape combinations}
\end{equation}
These quantities set the line spacing, the Bessel arguments and the relative weighting of the two polarisation amplitudes, respectively. For a narrow distribution of field inclinations, it is also an option to Taylor-expand in $\eta$ while retaining the Legendre expansion in $\mu$.

\subsection{Spectral basis via ultra-relativistic continuum kernels}
\label{sec: joint relativistic}

\begin{figure}[tbp]
\centering
    \includegraphics[width=\textwidth]{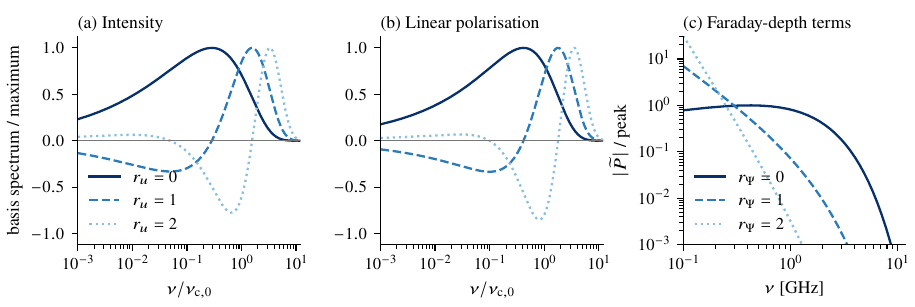}
    \caption{Continuum basis spectra of one cell for narrow channels. (a,\,b)~$F$ and $G$ with their first two derivatives in $u$, each divided by its peak. (c)~Polarisation basis spectra of the Faraday-depth terms for $\nu_{\rm c,0}=1\,$GHz and $|\Delta\Psi|=0.5\,\mathrm{rad\,m^{-2}}$, divided by the order-zero peak.}
    \label{fig: basis spectra}
\end{figure}

The continuum kernels of Eq.~\eqref{eq: directional continuum} depend on $(\gamma,B,\eta)$ through the characteristic frequency $\nu_{\rm c}$ and the amplitude factor $B_\perp$. We use a reference field strength $B_0>0$ to normalise the amplitude and put $B_\perp/B_0$ in the population weights. Elements with $B_\perp=0$ have zero weight in the continuum average. For $B_\perp>0$, with $\mathcal K_P^{\rm cont}\equiv\mathcal K_Q^{\rm cont}$ in the natural basis, the remaining channel kernel is
\begin{equation}
    \mathcal K^{\rm cont}_{Xj}(\nu_{\rm c},\Psi)
    \equiv\frac{B_0}{B_\perp}\int R_j(\nu)\,\mathcal K^{\rm cont}_X(\nu,\gamma,B,\eta)\,
    e^{i\mathsf s_X(c/\nu)^2\Psi}\diff\nu,
    \qquad X\in\{I,P\}.
    \label{eq: continuum channel kernel}
\end{equation}
When the pitch-angle density varies little across the beaming cone at fixed values of the other parameters, including the Faraday depth, we use Eq.~\eqref{eq: beaming limit} at each Faraday depth to carry out the pitch-angle integral. Grouping the contributions by characteristic frequency and Faraday depth gives
\begin{equation}
    \begin{aligned}
    X_j^{\rm obs}
    &\simeq\mathcal N_{\rm src}\int
    \mathcal W^{{\rm cont}(\mathsf s_X)}\,\mathcal K^{\rm cont}_{Xj}\,
    \diff\nu_{\rm c}\,\diff\Psi,\\
    \mathcal W^{{\rm cont}(\mathsf s_X)}(\nu_{\rm c},\Psi)
    &\equiv\int\mathcal P_{\rm F}(\boldsymbol p,\phi,\Psi)\,2\delta(\mu-\eta)\,\frac{B_\perp}{B_0}\,
    e^{i\mathsf s_X\phi}\,\delta\!\left[\nu_{\rm c}-\nu_{\rm c}(\boldsymbol p)\right]\,\diff\boldsymbol p\,\diff\phi.
    \end{aligned}
    \label{eq: reduced continuum}
\end{equation}
We Taylor-expand the channel kernel about the reference values $\nu_{\rm c,0}$ and $\Psi_{\rm ref}$, using $u=\ln(\nu_{\rm c}/\nu_{\rm c,0})$ and $\Delta\Psi=\Psi-\Psi_{\rm ref}$. As in Sec.~\ref{sec: joint general}, we write $\boldsymbol r=(r_u,r_\Psi)$, $\Delta^{\boldsymbol r}=u^{r_u}(\Delta\Psi)^{r_\Psi}$ and $\partial^{\boldsymbol r}=\partial_u^{r_u}\partial_\Psi^{r_\Psi}$. Keeping terms up to total degree $N$ gives
\begin{equation}
    \begin{aligned}
    X_j^{\rm obs}
    &\simeq\mathcal N_{\rm src}\sum_{|\boldsymbol r|\leq N}
    \widetilde X^{\rm cont}_{j\boldsymbol r}\,M^{{\rm cont}(\mathsf s_X)}_{\boldsymbol r},
    \qquad
    \widetilde X^{\rm cont}_{j\boldsymbol r}
    =\left.\frac{\partial^{\boldsymbol r}\mathcal K^{\rm cont}_{Xj}}{\boldsymbol r!}\right|_{u=0,\,\Psi_{\rm ref}},\\
    M^{{\rm cont}(\mathsf s_X)}_{\boldsymbol r}
    &\equiv\int\mathcal W^{{\rm cont}(\mathsf s_X)}\,\Delta^{\boldsymbol r}\,
    \diff\nu_{\rm c}\,\diff\Psi
    =\left\langle2\delta(\mu-\eta)\,\frac{B_\perp}{B_0}\,e^{i\mathsf s_X\phi}\,\Delta^{\boldsymbol r}\right\rangle_{\rm F}.
    \end{aligned}
    \label{eq: continuum joint response}
\end{equation}
At $N=2$, each cell has three real intensity moments and six complex polarisation moments. The zeroth intensity moment, multiplied by $\mathcal N_{\rm src}$, gives the electron column weighted by the pitch-angle density relative to isotropy and by $B_\perp/B_0$. Both weights are unity for isotropic pitch angles at $B_\perp=B_0$.
Figure~\ref{fig: basis spectra} illustrates the basis for narrow channels. Changing the reference characteristic frequency shifts the unrotated basis spectra along $\ln\nu$. The Faraday-depth terms matter most at low frequencies, where a given change in $\Psi$ rotates the polarisation more.

\paragraph{Field-strength degeneracy and asymptotic behaviour.}
The field strength enters the continuum kernels only through the amplitude $\propto B_\perp$ and $\nu_{\rm c}\propto B_\perp\gamma^2$ (Eq.~\eqref{eq: directional continuum}). A field-strength distribution can therefore be absorbed into an effective electron energy distribution if pitch angle, field direction and Faraday depth are independent of energy and field strength, provided the energy range covers the rescaled energies. Changing the assumed field strength in the effective model then rescales the inferred energies and electron column while giving the same predicted spectrum.
For $\nu_{\rm c}\gg\nu$, $F$ and $G$ follow the same power law, $(\nu/\nu_{\rm c})^{1/3}$. In this limit, a cell contributes only one moment combination per Faraday-depth order. For $\nu_{\rm c}\ll\nu$, a cell's contribution is exponentially suppressed.

\subsection{Coarse-graining and specified conditions}
\label{sec: coarse graining}

The expansions of Secs.~\ref{sec: joint general} and \ref{sec: joint relativistic} are local, each about a single reference point. Section~\ref{sec: cell lattice} combines such expansions over a lattice of cells to cover a broad ensemble. The cell moments are constrained by the non-negativity of the distribution and optionally by any specified form of it (Sec.~\ref{sec: special cases}). Statistical independence further makes the moments factorise, which reduces the number of free coefficients.

\subsubsection{The lattice of moments}
\label{sec: cell lattice}
We divide a broad ensemble into cells in the expansion variables and
expand about a separate reference point in each cell. For linear
polarisation, the harmonic construction (Sec.~\ref{sec: joint general}) uses a lattice in $(\gamma,B,\Psi)$, and the continuum construction
(Sec.~\ref{sec: joint relativistic}) uses a lattice in $(\nu_{\rm c},\Psi)$.
Figure~\ref{fig: framework overview} shows these lattices, with each
arrow representing the moment vector of one cell. For Stokes components
independent of $\Psi$, the lattice has one fewer dimension.

For either kernel, \textsc{SyncEM} supports both physical and reduced
parameterisations. The physical parameterisation is convenient when
specifying a pitch-angle distribution or a model for the electron energy
spectrum, as in Sec.~\ref{sec: classical models}.

With the integrals of Eqs.~\eqref{eq: finite joint response} and \eqref{eq: continuum joint response} restricted to cell $b$, the channel prediction is
\begin{equation}
    X_j^{\rm obs}\simeq\mathcal N_{\rm src}\sum_b\widetilde{\boldsymbol X}_{jb}\cdot\boldsymbol M_b,
    \qquad
    \boldsymbol M_b=\bigl\{M^{(\mathsf s_X)}_{\ell k\boldsymbol r}\bigr\}_b
    \ \ \text{or}\ \
    \bigl\{M^{{\rm cont}(\mathsf s_X)}_{\boldsymbol r}\bigr\}_b .
    \label{eq: cell sum}
\end{equation}
The vector $\boldsymbol M_b$ collects the moments of cell $b$ and $\widetilde{\boldsymbol X}_{jb}$ the corresponding basis spectra. When both kernels are used, the low-energy part of the ensemble enters Eq.~\eqref{eq: reduced general} and the high-energy part Eq.~\eqref{eq: reduced continuum}, and the two sets of cells add. Each $\boldsymbol M_b$ is a coarse-grained summary of the reduced distribution in its cell. Without a specified population model, the fit can treat the cell moments as separate coefficients.
Suitable orders and reference points depend on the channel width and the parameter range. The examples of Sec.~\ref{sec: classical models} specify the field, angle and Faraday-depth statistics and use cells in $\ln\gamma$ (Appendix~\ref{app: test ensembles}).
\paragraph{Choice of reference points.}
For SED fitting, the reference points can be fixed in advance or fitted together with the moments:
\begin{enumerate}
    \item \textbf{Fixed reference points:} The basis is computed once, and the fit of the cell moments in Sec.~\ref{sec: linear fit} is linear and solved in closed form. \textsc{SyncEM} builds the basis at reference points chosen by the user, and the examples of Sec.~\ref{sec: classical models} fix the energy reference points at the cell centres.
    \item \textbf{Fitted reference points:} A reference point placed at the mean of the expansion variable in its cell, as Ref.~\cite{chluba2017rethinking} does for the pivot index of the $\beta$-moment SED, automatically sets the first moment of that variable to zero. The basis then depends on the fit parameters, so the fit becomes nonlinear and the basis has to be recomputed for each trial reference point.
\end{enumerate}
Section~\ref{sec: classical ensembles} measures the effect of the order and of the reference-point placement in the energy cells for the example populations. Measurements on a finite set of populations give no general convergence guarantee. Iterative approaches for improving the representation by selecting better pivots with the goal to reduce the number of moments that are required to obtain an accurate model could remedy these issues but are left to future exploration.

\subsubsection{Constrained moments}
\label{sec: special cases}

\paragraph{Intrinsic constraints.}
The intensity distributions are non-negative, so not every set of moment values in a cell is admissible. A fit can impose this condition without a model of the distribution. The treatment used in Sec.~\ref{sec: cmb extrapolation}, non-negative cells, represents the distribution in each cell by non-negative weights at fixed energies and adds a penalty that damps the combinations the data constrain poorly (Appendix~\ref{app: test ensembles}).

\paragraph{Uncoupled angular statistics.}
\label{sec: uncoupled angles}

If the angular pair $(\mu,\eta)$ is independent of $(\gamma,B,\phi,\Psi)$, the joint moments factorise as
\begin{equation}
    M^{(\mathsf s_X)}_{\ell k\boldsymbol r}=\Theta_{\ell k}\,
    \left\langle e^{i\mathsf s_X\phi}\Delta^{\boldsymbol r}\right\rangle_{\rm F},
    \qquad \Theta_{\ell k}\equiv\langle\Leg_\ell(\mu)\Leg_k(\eta)\rangle_{\rm F},
    \label{eq: uncoupled moments}
\end{equation}
where the $\Theta_{\ell k}$ are the Legendre coefficients of the angular distribution. For retained angular degrees $L_\mu,L_\eta$, the angular statistics then enter through at most $(L_\mu+1)(L_\eta+1)$ numbers shared by all energy, field and Faraday-depth moments. With known $\Theta_{\ell k}$ the prediction remains linear in the non-angular moments. With unknown $\Theta_{\ell k}$ it is bilinear, a sum of products of $\Theta_{\ell k}$ and non-angular moments, so the linear fit of Sec.~\ref{sec: linear fit} determines either set only with the other held fixed.
%

\paragraph{Independent Faraday-screen limit.}
\label{sec: independent screen limit}

\begin{figure}[tbp]
\centering
\includegraphics[width=\textwidth]{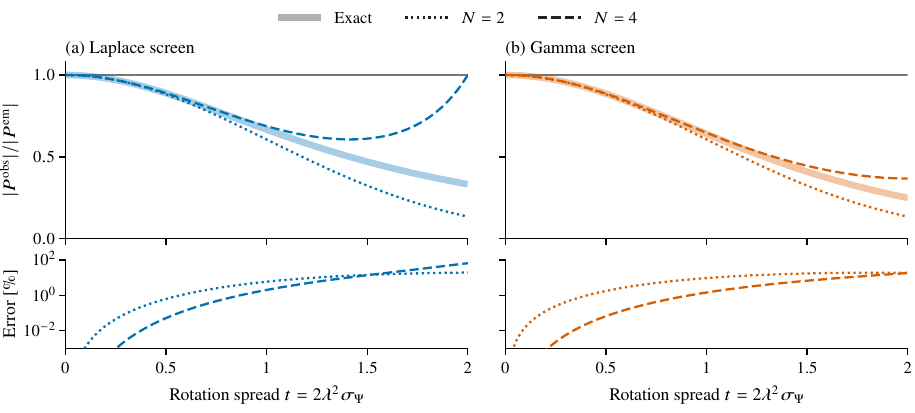}
\\
\caption{Cumulant truncation of $C_{\rm scr}$ for (a)~a Laplace and (b)~a skewed Gamma Faraday-depth distribution with the same mean and variance. Upper panels: remaining polarisation fraction. Lower panels: error of $\widehat C_N$ in per cent of $|P^{\rm em}|$.}
\label{fig: screen shapes}
\vspace{-2mm}
\end{figure}

For Faraday depth independent of the emitting parameters, the joint density factorises as $\mathcal P_{\rm F}(\boldsymbol p,\phi,\Psi)=\mathcal P_{\rm col}(\boldsymbol p,\phi)\mathcal P_{\rm scr}(\Psi)$.
We call this distribution of Faraday depth, independent of the emission, the screen. When emission and rotation are mixed, the Faraday depth varies along the ray, and emitting parameters that also vary along the ray are in general correlated with it. Zero covariance between Faraday depth and each emitting parameter is not sufficient for independence, because it excludes only linear dependence on one parameter at a time. For a specified screen, we average the Faraday phase directly without a Taylor expansion in Faraday depth. Let $P_\nu^{\rm em}=\mathcal N_{\rm src}\langle e^{2i\phi}\mathcal K_P^{\rm nat}(\nu)\rangle_{\rm col}$ be the polarisation in the absence of Faraday rotation. With $\tau=2\lambda^2$ at fixed $\nu>0$ and $\lambda=c/\nu$ in metres, the full Faraday-depth average is
\begin{equation}
    P_\nu^{\rm obs}=P_\nu^{\rm em}C_{\rm scr}(\tau),
    \qquad C_{\rm scr}(\tau)=\langle e^{i\tau\Psi}\rangle_{\rm scr}.
    \label{eq: independent screen response}
\end{equation}
Here $C_{\rm scr}$ is the characteristic function of the Faraday-depth distribution, and $\ln C_{\rm scr}$ is its cumulant generating function. For a channel this becomes
\begin{equation}
 P_j^{\rm obs}=\int R_j(\nu)P_\nu^{\rm em}
 C_{\rm scr}\!\left(2c^2/\nu^2\right)\diff\nu.
 \label{eq: channel screen response}
\end{equation}
Each harmonic line carries $C_{\rm scr}$ at its own frequency, and a single factor at the channel centre is a further approximation. With Faraday-depth cumulants $\kappa_n$, where $\kappa_1=\bar\Psi$ and $\kappa_2=\sigma_\Psi^2$, truncating $\ln C_{\rm scr}$ at order $N$ gives
\begin{equation}
 \widehat C_N(\tau)=\exp\Bigl[\sum_{n=1}^{N}\frac{(i\tau)^n}{n!}\,\kappa_n\Bigr]
 \label{eq: screen cumulant truncation}
\end{equation}
near $\tau=0$, where a continuous logarithm of $C_{\rm scr}$ exists. 
A Gaussian distribution of Faraday depth has no cumulants above second order, so $\widehat C_2=\exp(i\tau\bar\Psi-\tau^2\sigma_\Psi^2/2)$ is exact for any value of $\tau\sigma_\Psi$~\cite{burn1966depolarization}. For other distributions the error of $\widehat C_N$ depends on their shapes and on the rotation spread $\tau\sigma_\Psi$, the standard deviation of the Faraday phase (Fig.~\ref{fig: screen shapes}, for a Laplace and a skewed Gamma distribution with the same mean and variance as the Gaussian). If a retained cumulant above second order is nonzero, $\widehat C_N$ is not the characteristic function of any distribution~\cite{marcinkiewicz1939gauss}. More generally, when Faraday depth is correlated with the emitting parameters, the Taylor expansion in $\Delta\Psi$ can be replaced by a cumulant expansion of the conditional Faraday-depth distribution at fixed emitting parameters.

\paragraph{Mixtures of power-law electrons.}
For a mixture of power laws on the fixed energy range, with index density $g(p)$, pivot $\gamma_{\rm piv}$ and $\zeta=\ln(\gamma/\gamma_{\rm piv})$,
\begin{equation}
    \frac{\diff\mathcal N}{\diff\gamma}=A\int g(p)\Bigl(\frac{\gamma}{\gamma_{\rm piv}}\Bigr)^{-p}\diff p
    =A\,\bigl\langle e^{-p\zeta}\bigr\rangle_g .
    \label{eq: index mixture}
\end{equation}
The average is the moment-generating function of $p$ evaluated at $-\zeta$. It has two expansions.
The moment expansion about an index $p_0$ is
\begin{equation}
 \frac{\diff\mathcal N}{\diff\gamma}=A\,e^{-p_0\zeta}\sum_{n=0}^{\infty}\frac{m_n(-\zeta)^n}{n!},\qquad
 m_n=\bigl\langle(p-p_0)^n\bigr\rangle_g ,
 \label{eq: index moment expansion}
\end{equation}
which is linear in $Am_n$. Each term therefore gives one basis spectrum $\int\overline{\mathcal K}_{Xj}(\gamma)\,e^{-p_0\zeta}(-\zeta)^n/n!\,\diff\gamma$ over the energy range, where $\overline{\mathcal K}_{Xj}(\gamma)$ is the channel kernel averaged over the specified statistics of the other variables (Appendix~\ref{app: test ensembles}). Each basis spectrum is computed once and fitted as in Sec.~\ref{sec: linear fit}. This expansion is the electron counterpart of the expansion of Ref.~\cite{chluba2017rethinking}, which mixes spectral indices of the emission in $\ln\nu$. Truncation at order $N$ omits all higher moments, so it requires the index spread times $|\zeta|$ to be small over the energy range.
The cumulant expansion is
\begin{equation}
 \frac{\diff\mathcal N}{\diff\gamma}=A\exp\Bigl[\sum_{n=1}^{\infty}\frac{k_n(-\zeta)^n}{n!}\Bigr],\qquad
 k_1=\bar p,\quad k_2=\sigma_p^2,
 \label{eq: index cumulant expansion}
\end{equation}
with $k_n$ the cumulants of $g$. 
At the same truncation order, the cumulant form retains more of the statistics of $g$. It sets only the cumulants above order $N$ to zero, while the moment form drops the higher moments, which also contain products of lower cumulants. 
However, when its truncation error is negligible, the moment expansion is preferable because it provides a linear representation and the fit is solved in closed form.

The test ensembles E1 and E3 (Sec.~\ref{sec: classical models}) are the single-index and Gaussian cases of Eq.~\eqref{eq: index mixture}. A single power law, $g\propto\delta(p-p_0)$, has $m_n=0$ for $n\geq1$ and $k_n=0$ for $n\geq2$. A Gaussian $g$ has $k_n=0$ for $n\geq3$, so the second-order cumulant form is exact for any $\sigma_p$. 
For the Gaussian mixture, the moment expansion about $p_0=\bar p$ has $m_{2n}=(2n-1)!!\,\sigma_p^{2n}$ and converges slowly once $\sigma_p|\zeta|\sim1$ ($1.1$ for E3 at the upper energy limit).

\subsection{Linear SED fitting}
\label{sec: linear fit}

For SED fitting, we compress the linear representation of the moment expansion into a more compact one. The basis can contain many more moments than the measured channels constrain, so we retain only the moment combinations to which the channels respond.

We denote the measured channel Stokes components as $\boldsymbol d$ and the real components of the unknown moments, multiplied by $\mathcal N_{\rm src}$, as $\boldsymbol a$. Each moment is expressed in units of a fixed scale chosen before the fit (Eq.~\eqref{eq: fit coefficient scales} for the examples), so that the entries of $\boldsymbol a$ are dimensionless and of comparable size. The real design matrix $\boldsymbol C$ holds the basis spectra in the same units, assembled from $\widetilde I$, $\widetilde V$ and the real and imaginary parts of $\widetilde P$, so that $\boldsymbol C\boldsymbol a$ is the predicted spectrum.
We model $\boldsymbol d\simeq\boldsymbol C\boldsymbol a+\boldsymbol\epsilon$, with zero-mean Gaussian noise $\boldsymbol\epsilon$ of fixed covariance $\boldsymbol\Sigma$. The singular-value decomposition of the noise-weighted basis and the estimator retaining $R$ singular directions are
\begin{equation}
    \boldsymbol\Sigma^{-1/2}\boldsymbol C=\boldsymbol U\boldsymbol D\boldsymbol V^{\mathsf T},\qquad
    \boldsymbol\beta=\boldsymbol V_R^{\mathsf T}\boldsymbol a,\qquad
    \widehat{\boldsymbol\beta}=\boldsymbol D_R^{-1}\boldsymbol U_R^{\mathsf T}\boldsymbol\Sigma^{-1/2}\boldsymbol d.
    \label{eq: svd estimator}
\end{equation}
We choose $R$ by the unbiased predictive-risk estimate (UPRE \cite{vogel2002computational}), $R_\star=\arg\min_R[\chi^2(R)+2R]$, where $\chi^2(R)$ is the noise-weighted squared residual of the fitted channels. Setting the components along the discarded singular directions to zero gives the minimum-norm $\boldsymbol a$ consistent with $\widehat{\boldsymbol\beta}$. These zeros are not inferred physical values.

A physical model with parameters $\boldsymbol\theta$ predicts moments $\boldsymbol a(\boldsymbol\theta)$ as low-dimensional integrals of its distribution. We fit its parameters to the retained combinations by minimising the compressed $\chi^2$
\begin{equation}
 \chi^2_\beta(\boldsymbol\theta)=\left[\widehat{\boldsymbol\beta}-\boldsymbol V_R^{\mathsf T}\boldsymbol a(\boldsymbol\theta)\right]^{\mathsf T}\boldsymbol D_R^2\left[\widehat{\boldsymbol\beta}-\boldsymbol V_R^{\mathsf T}\boldsymbol a(\boldsymbol\theta)\right].
 \label{eq: compressed likelihood}
\end{equation}
The full $\chi^2$, $|\boldsymbol\Sigma^{-1/2}[\boldsymbol d-\boldsymbol C\boldsymbol a(\boldsymbol\theta)]|^2$, serves as a reference, with which Sec.~\ref{sec: moment combinations} compares the compressed fits of the examples.

\FloatBarrier

\section{Examples and physical interpretation}
\label{sec: classical models}
We test forward predictions from specified populations, fits to channel data that prescribe no electron-spectrum family, and extrapolation up to $300\,\mathrm{GHz}$. The electron spectra are a power law (E1), a smoothly broken power law (E2), an index mixture (E3), and aged spectra with isotropised (E4-JP, the Jaffe--Perola model~\cite{jaffe1973dynamical}) or fixed (E4-KP, the Kardashev--Pacholczyk model~\cite{kardashev1962nonstationarity,pacholczyk1970radio}) pitch angles. E5 (two power-law zones) and E6 (ageing with losses in the local field) correlate electron energy with field azimuth and with field strength, respectively. E1M places the E1 electrons in a Maxwellian field-strength distribution. Appendix~\ref{app: test ensembles} defines the ensembles (Table~\ref{tab: classical ensembles}) and gives their numerical settings. The electron energies span $100\leq\gamma\leq2.96\times10^6$. For intensity and linear polarisation we use the harmonic kernels below the switch at $\gamma=300$ and the continuum kernels of Sec.~\ref{sec: continuum limit} above it. At these field strengths the electrons below the switch add less than $10^{-10}$ of the intensity in every channel.

\subsection{Predictions from population moments}
\label{sec: classical ensembles}
The basis uses 41 channels over $0.1$--$10\,\mathrm{GHz}$. We divide $\ln\gamma$ into nineteen cells above the switch and two below it, and expand to order $N$ in each cell (Appendix~\ref{app: test ensembles}), with $N=2$ in all fits. In the continuum limit, variations of field strength are degenerate with the electron energy distribution under the independence assumptions of Sec.~\ref{sec: joint relativistic}, so the SED alone does not separate them. The tests therefore specify the field statistics and an independent Faraday screen when the basis is built. A population then determines its moments, and the moments weight the basis spectra to predict the channel spectrum. Integrating the same population directly against the same kernels gives the true spectrum, and its difference from the prediction is the representation error of the basis. For E2 with the continuum kernels, on one processor core, a direct integration of the channel spectrum takes $1.8\,$s, the 57 basis spectra take $0.15\,$s to build, and each prediction from the moments then takes $3\times10^{-5}\,$s.
With the reference points at the cell centres, the largest prediction error of E1--E4-JP over the channels and $I,Q,U$ falls from $2.6$--$18\%$ of $I_j$ at $N=0$ to $0.086$--$0.36\%$ at $N=2$ and $0.0088$--$0.045\%$ at $N=4$ (Table~\ref{tab: classical fit summary}). Shifting the reference points by a quarter of a cell width to lower or higher energy raises the intensity error of E1, E2 and E4-JP to at most $6.4\%$ at $N=2$ and $0.92\%$ at $N=4$.

\begin{table}[tbp]
\centering
\footnotesize
\setlength{\tabcolsep}{3pt}
\caption{Predictions at the true moments and fits with $1\%$ noise. Errors are the largest deviation from the true spectrum over the 41 channels and $I,Q,U$, in per cent of $I_j$. $R_\star$ and $\chi^2_{\rm fit}$ are for the noise realisation of Fig.~\ref{fig: classical fits}, and the fit error is the median over 4000 realisations (400 for the $^\ddagger$ row). E1M uses the basis built for $B\sim U[4,6]\,\mu$G and is predicted from the E1 moments. $^\dagger$Predicted from the moments of the JP spectrum that matches E4-KP (Appendix~\ref{app: test ensembles}). $^\ddagger$Azimuth assumed independent of energy, with one complex azimuth factor shared by all cells. Its real and imaginary parts are the $+2$ in the $R_\star$ column.}
\label{tab: classical fit summary}
\begin{tabular}{@{}llccccc@{}}
\toprule
 & & \multicolumn{2}{c}{Prediction error [\%]} & \multicolumn{3}{c}{Fit at $N=2$} \\
\cmidrule(lr){3-4}\cmidrule(lr){5-7}
Ensemble & Case & $N=2$ & $N=4$ & $R_\star$ & $\chi^2_{\rm fit}$/dof & Median error [\%] \\
\midrule
E1 & power law & 0.086 & 0.0088 & 7 & 52.1/56 & 1.5 \\
E2 & smoothly broken power law & 0.091 & 0.012 & 7 & 49.9/56 & 1.4 \\
E3 & index mixture & 0.10 & 0.0096 & 7 & 52.5/56 & 1.5 \\
E4-JP & ageing, isotropised pitch & 0.36 & 0.045 & 9 & 49.0/54 & 1.6 \\
E4-KP$^\dagger$ & ageing, fixed pitch & 0.89 & 0.064 & 9 & 48.4/54 & 1.6 \\
E5 & two zones, joint moments & 0.18 & 0.014 & 17 & 43.2/46 & 1.7 \\
E1M & Maxwellian field & 3.8 & 3.8 & 7 & 52.1/56 & 1.5 \\
E5$^\ddagger$ & two zones, one azimuth factor & 29 & 29 & 7+2 & 270.9/54 & 6.1 \\
\bottomrule
\end{tabular}
\end{table}

Ignoring the correlation of electron energy with the field gives large errors in E5 and E6, and including it in the basis brings them below $1\%$ at $N=2$. For E5, assuming the azimuth to be independent of energy gives an error of $29\%$ at both $N=2$ and $N=4$, so raising the order does not reduce it. The joint moments of energy and azimuth in each cell (Eq.~\eqref{eq: finite joint response}) reduce it to $0.18\%$. For E6, basis spectra averaged over the marginal field-strength distribution give an error of $8.4\%$, and the joint energy--field basis (Appendix~\ref{app: test ensembles}) gives $0.78\%$.

\subsection{Spectral fits and comparisons}
\label{sec: sed reconstruction}

\begin{figure}[tbp]
\centering
\includegraphics[width=\textwidth]{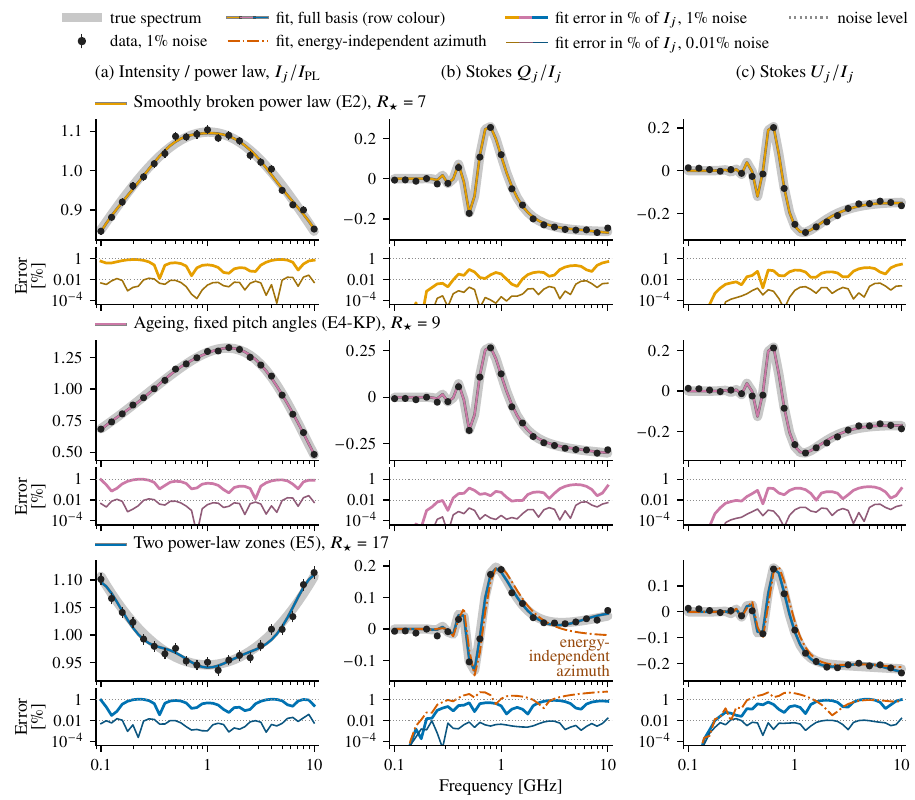}
\caption{Fits of the $N=2$ basis to 21 of the 41 channels for E2, E4-KP and E5 (rows). $I_{\rm PL}$ is a power law fitted to the true spectrum. Lower panels: errors in all 41 channels, including the 20 left out of the fit. The errors of the full-basis fits are close to the noise level. The E5 fit with an energy-independent azimuth deviates from the true $Q$ and $U$ by up to $6\%$ of $I_j$ and is rejected by its $\chi^2$.}
\label{fig: classical fits}
\end{figure}

For the fits we add independent Gaussian noise with standard deviation $0.01I_j$ to each of $I,Q,U$, fit every second channel (21 of the 41) and compare the predictions with the true spectrum in all 41 channels, including the 20 left out. The rank $R_\star$ is selected by UPRE (Sec.~\ref{sec: linear fit}). Figure~\ref{fig: classical fits} shows three of these fits and Table~\ref{tab: classical fit summary} lists all of them.

\paragraph{The moment-expansion basis fits the emission of every tested electron energy spectrum to within the noise.}
For the noise realisation of Fig.~\ref{fig: classical fits} and Table~\ref{tab: classical fit summary}, every fit except the E5 fit with an energy-independent azimuth (a wrong assumption by design) has $\chi^2$ consistent with its degrees of freedom. For each of these seven fits, the median over 4000 noise realisations of the largest error over the 41 channels is $1.4$--$1.7\%$, close to the $1\%$ noise.
At $0.01\%$ noise the $I,Q,U$ fits of Fig.~\ref{fig: classical fits} retain more combinations, which describe finer spectral detail, and their errors fall below $0.04\%$ of the intensity (lower panels).

\paragraph{The $\chi^2$ test rejects the fit that deliberately omits the energy--azimuth correlation of E5.}
We also fit E5 (two zones with different electron indices and field directions) under the wrong assumption that the field azimuth is independent of electron energy. This fit does not match $Q$ and $U$ and is rejected by its $\chi^2$. The fit with joint moments of energy and azimuth matches $Q$ and $U$ within the noise.

\paragraph{Breaks and cut-offs in the electron spectrum limit the accuracy of empirical spectral laws, however low the noise.}
Empirical spectral laws are widely used to model the intensity, so we compare them with the basis on $I$ alone (Table~\ref{tab: classical baselines}). At $1\%$ noise the basis predicts the intensities of the left-out channels to better than $1\%$ for E1--E5. The best three- or four-parameter law also reaches this accuracy for E1--E3 and E5, whose spectra are close to its form. The aged spectra represent spectra that steepen towards a cut-off set by energy losses, and most of these laws do not fit them. Only the four-parameter $\beta$-moment SED of Ref.~\cite{chluba2017rethinking} with $\omega_1$ free reaches this accuracy for them, and the power-law and curved-power-law forms of \textsc{PySM}~\cite{thorne2017pysm,zonca2021pysm3} are off by more than $5\%$.
At lower noise the error of the basis keeps falling, while that of a law falls only until it reaches a floor set by its form. For E1 and E3 the best law reaches no floor down to $0.01\%$ noise and is more accurate than the basis. For E2, E5 and the aged spectra the floor of the best law lies above the basis error at $0.01\%$ noise. The same holds for a spectrum computed with \textsc{GALPROP}~\cite{orlando2013synchrotron}, whose field model, kernel and electron spectrum were not used to build the basis (Appendix~\ref{app: test ensembles}).

\subsection{\texorpdfstring{Extrapolation up to 300\,GHz}{Extrapolation up to 300 GHz}}
\label{sec: cmb extrapolation}
\begin{figure}[tbp]
\centering
\includegraphics[width=\textwidth]{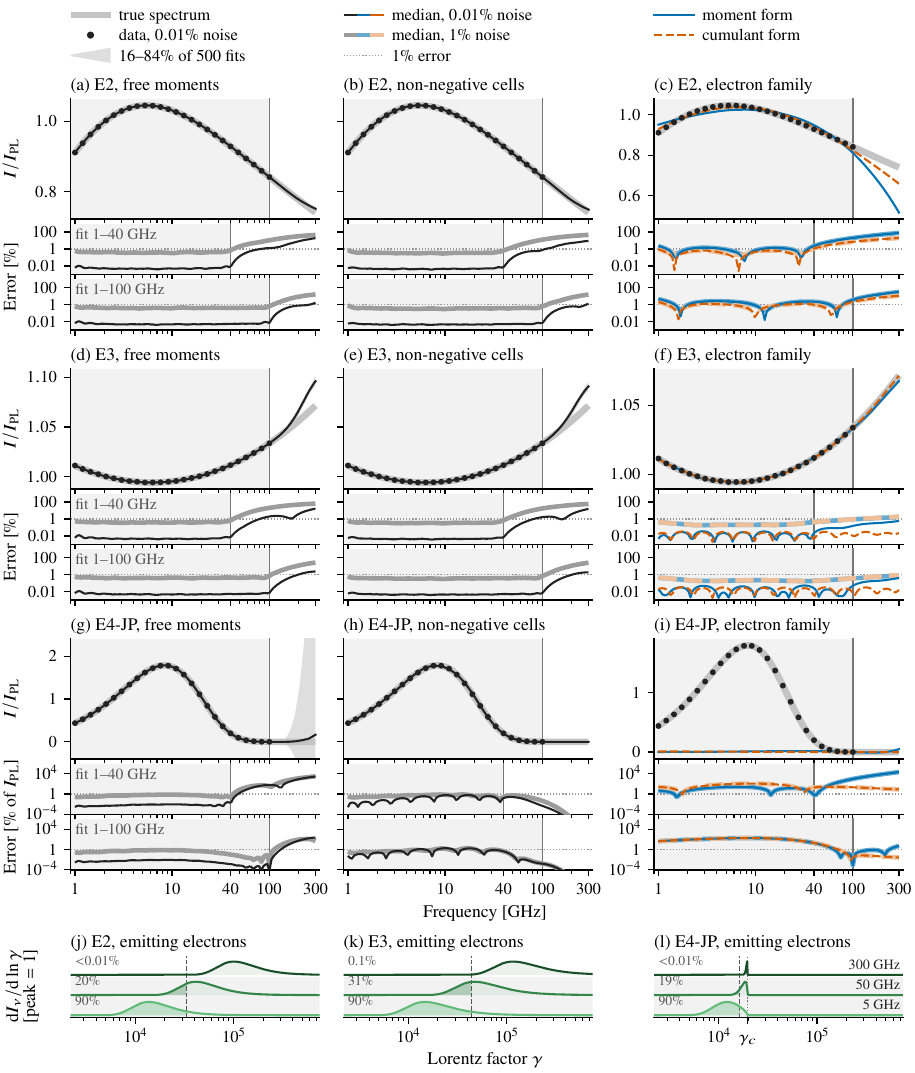}
\caption{Fits to E2, E3 and E4-JP (rows) and their extrapolations up to $300\,$GHz with three treatments (columns). Upper parts: intensities divided by a power law fitted to the truth over $1$--$40\,$GHz, for fits over $1$--$100\,$GHz (shaded) to data with $0.01\%$ noise. Lower parts: median errors in units of the intensity for E2 and E3 and of the power law for E4-JP, whose emission above its cut-off is negligible. Free moments do not constrain the electrons that emit mainly above the band. Non-negative cells bound them where the spectrum is cut off, and an electron family fixes them through its parameters. 
(j)--(l): intensity per unit $\ln\gamma$ at $5$, $50$ and $300\,$GHz (bottom to top), scaled to unit peak. The numbers give the share from electrons left of the dashed line.}
\label{fig: cmb extrapolation}
\end{figure}

We fit the $I$, $Q$ and $U$ data of E2, E3 and E4-JP over $1$--$40$ or $1$--$100\,\mathrm{GHz}$, with noise of $1\%$ or $0.01\%$ of the true intensity in each channel, and compare the predicted intensity with the truth up to $300\,\mathrm{GHz}$ (Fig.~\ref{fig: cmb extrapolation}). Appendix~\ref{app: test ensembles} gives the setup, and we quote median errors over 500 noise realisations. The fitted band constrains only the moment combinations that shape the spectrum in that band. Part of the emission at higher frequencies depends on other combinations, which involve mainly the moments of the high-energy electrons. Figure~\ref{fig: cmb extrapolation}(j)--(l) shows which electrons emit at each frequency. The electrons left of the dashed lines produce $90\%$ of the $5\,\mathrm{GHz}$ intensity, $19$--$31\%$ of the $50\,\mathrm{GHz}$ intensity and at most $0.1\%$ of the $300\,\mathrm{GHz}$ intensity. E4-JP has no electrons above its cut-off $\gamma_c$, and electrons within $5\%$ of $\gamma_c$ produce $88\%$ of its $300\,\mathrm{GHz}$ intensity. We compare three treatments of these electrons. 
All three hold the field, pitch-angle and Faraday-screen statistics of the ensembles at their true values, so only the electron energy distribution is fitted.

Free moments are the moments of the energy cells, with no relation imposed between cells and no electron-spectrum family, and the fit of Sec.~\ref{sec: linear fit} sets to zero the combinations that the band does not constrain. The second treatment, non-negative cells, requires the electron distribution in each cell to be non-negative and damps the combinations that the band constrains poorly. The third is an electron family, the power-law mixture of Eq.~\eqref{eq: index mixture} truncated at second order, in its moment form about $p_0=2.6$ and in its cumulant form. All fits use the $N=2$ basis, for which non-negative cells give a non-negative intensity over $1$--$300\,\mathrm{GHz}$. At higher order the polynomial approximation of a cell's kernel can be negative, so non-negative cells no longer ensure a non-negative intensity.

\paragraph{Free moments need no model of the electron spectrum, while an electron family is accurate only for spectra it contains.}
Fitted over $1$--$100\,\mathrm{GHz}$, free moments predict the $300\,\mathrm{GHz}$ intensity of E2 with an error of $16\%$ at $1\%$ noise and $1.6\%$ at $0.01\%$ noise, and that of E3 with $26\%$ and $2.3\%$. At $0.01\%$ noise this error is mainly a bias. An electron family fixes the moments at all energies with a few parameters. The extrapolation of E3 by an electron family is accurate (Fig.~\ref{fig: cmb extrapolation}f) because E3 is a Gaussian mixture of power laws (Eq.~\eqref{eq: index mixture}), which the cumulant form contains exactly and the moment form approximates closely. Fitted over $1$--$100\,\mathrm{GHz}$, the cumulant form predicts the $300\,\mathrm{GHz}$ intensity to $0.8\%$ at $1\%$ noise and $0.015\%$ at $0.01\%$ noise, the limit set by the $N=2$ basis. Fitted over $1$--$40\,\mathrm{GHz}$ at $1\%$ noise, it predicts the $300\,\mathrm{GHz}$ intensity to $1.9\%$. Neither form follows the cut-off of E4-JP, and the in-band $\chi^2$ rejects all of these fits. E2 is a broken power law, which neither form contains. Both forms fit its $1$--$40\,\mathrm{GHz}$ data at $1\%$ noise in most realisations, yet their $300\,\mathrm{GHz}$ errors are $20\%$ (cumulant) and $78\%$ (moment) at both noise levels. An acceptable in-band fit therefore does not validate extrapolation to higher frequencies, which depends on the assumed electron spectrum at higher energies. The smoothly broken power-law family eSBPL (Appendix~\ref{app: test ensembles}), which contains E2, predicts its $300\,\mathrm{GHz}$ intensity to $0.9\%$ at $1\%$ noise and $0.08\%$ at $0.01\%$ noise when fitted over $1$--$100\,\mathrm{GHz}$, and to $2.6\%$ at $1\%$ noise over $1$--$40\,\mathrm{GHz}$.

\paragraph{Non-negative cells follow the cut-off of an aged spectrum and match or improve on free moments for smooth spectra.}
E4-JP has no electrons above its cut-off, and its intensity at $300\,\mathrm{GHz}$ is $6\times10^{-12}$ of the power law fitted over $1$--$40\,\mathrm{GHz}$. Free moments fitted over $1$--$100\,\mathrm{GHz}$ do not constrain the moments of the highest-energy cells, which are zero for E4-JP. The median error of the free-moment prediction at $300\,\mathrm{GHz}$ is $0.45$ times that power law at $1\%$ noise and $2.3$ times at $0.01\%$ noise. With non-negative cells, electrons above the cut-off would add emission at the highest fitted frequencies, which the data exclude, so the best fit leaves them out. For the same band at $1\%$ noise, the largest $300\,\mathrm{GHz}$ intensity among non-negative fits whose $\chi^2$ exceeds the minimum by at most one has a median of $2\times10^{-5}$ of the power law. The in-band $\chi^2$ accepts most non-negative cell fits to E4-JP at $1\%$ noise and rejects them at $0.01\%$ noise, because non-negative $N=2$ cells do not reproduce the spectrum near the cut-off to within this noise. For E2 and E3, where every cell holds electrons, the $300\,\mathrm{GHz}$ error with non-negative cells is similar to or smaller than that of free moments, $1.3\%$ against $1.6\%$ for E2 and $1.8\%$ against $2.3\%$ for E3 over $1$--$100\,\mathrm{GHz}$ at $0.01\%$ noise.

\paragraph{A wider band and lower noise improve the free-moment extrapolation of smooth spectra.}
Widening the band from $1$--$40$ to $1$--$100\,\mathrm{GHz}$ lowers the $300\,\mathrm{GHz}$ error of free moments for E2 from $45\%$ to $16\%$ at $1\%$ noise and from $19\%$ to $1.6\%$ at $0.01\%$ noise, and that of E3 from $65\%$ to $26\%$ and from $17\%$ to $2.3\%$. For E2 and E3 the polarisation error, divided by the true $|P|\simeq0.3I$, is of the same order as the intensity error: for free moments and non-negative cells fitted over $1$--$100\,\mathrm{GHz}$ it is $0.5$--$1.4\%$ of $I$ at $0.01\%$ noise and $3.7$--$5.7\%$ at $1\%$ noise.

\paragraph{An in-band fit within the noise does not validate an empirical extrapolation.}
We also fit the empirical laws of Table~\ref{tab: classical baselines} to the intensity in these bands, and fit the laws and free moments over $7.6$--$304\,\mathrm{GHz}$ to predict the $1\,\mathrm{GHz}$ intensity (Appendix~\ref{app: test ensembles}, Table~\ref{tab: law extrapolation}). 
For the broken power law at $1\%$ noise its in-band $\chi^2$ accepts most fits, yet it misses the target intensities by $11$--$23\%$. Free-moment fits over $7.6$--$304\,\mathrm{GHz}$ are accepted for E1--E3 and miss the $1\,\mathrm{GHz}$ intensity by more than half, because these data constrain the electrons that emit mainly at $1\,\mathrm{GHz}$ only weakly.

\subsection{Physical interpretation and model dependence}
\label{sec: moment combinations}

The fit returns the $R_\star$ combinations $\widehat{\boldsymbol\beta}$ of Eq.~\eqref{eq: svd estimator}. Physical parameters follow from them only through a model of the electron and field distributions.

\paragraph{An electron-spectrum model fitted to the retained combinations recovers its parameters when its assumptions hold.}
A parametric electron spectrum predicts the combinations through its moments, so its parameters can be fitted with Eq.~\eqref{eq: compressed likelihood} without repeating the radiation integrals. For E1--E4-JP all parameters in Table~\ref{tab: classical parameters} lie within one standard deviation of their true values, and their uncertainties agree with the scatter over noise realisations (Appendix~\ref{app: test ensembles}). These tests assume the field model and electron family used to generate the data.

\paragraph{A wrong field-strength distribution or ageing model shifts the inferred parameters without degrading the fit.}
With the field geometry and an independent Faraday screen held fixed, the continuum kernels of Eq.~\eqref{eq: directional continuum} allow a common rescaling of field strengths to be offset by rescaling the electron energies, density normalisation and assumed energy range. The spectrum then constrains only parameter combinations that this rescaling leaves unchanged, such as a break frequency $\nu_b\propto\gamma_c^2\langle B_\perp\rangle_{\rm col}$ and the amplitude $A\langle B_\perp^{(p+1)/2}\rangle_{\rm col}$, with $\gamma_c$ the cut-off Lorentz factor of an aged spectrum and $A$ the normalisation of $\gamma^{-p}$. Fitted with the basis built for $B\sim U[4,6]\,\mu$G and the E1 family, noise-free E1M data give a density amplitude at $\gamma=10^4$ that is $3.7\%$ low, $17.2$ standard deviations for the assumed $1\%$ noise, while $\chi^2$ stays at the value of the noise-free E1 fit. The amplitude $A\langle B_\perp^{(p+1)/2}\rangle_{\rm col}$ is recovered to $0.08\%$. Fitted with the JP family, E4-KP data give an age $15\%$ longer than the true age, and at $1\%$ noise the $\chi^2$ does not distinguish the two ageing models, because, for the pitch angles $|\mu|\leq1/2$ observed here and the inverse-Compton losses of the test, a JP spectrum with a longer age reproduces the KP spectrum to well within the noise over $0.1$--$10\,\mathrm{GHz}$ (Appendix~\ref{app: test ensembles}).

\paragraph{Comparing fits under different assumptions can exclude a model if the other assumptions hold.}
Of the three misspecified fits (E1M, E4-KP and E5 with an energy-independent azimuth), only the E5 fit is rejected by $\chi^2$ at $1\%$ noise, while the fit with joint moments of energy and azimuth is accepted (Sec.~\ref{sec: sed reconstruction}). Both fits hold the field statistics and the independent screen at their true values, and the rejection indicates a correlation of energy and azimuth only if these assumptions hold. A polarisation angle that changes with frequency can also come from a correlation of emission with Faraday depth or from a misspecified screen, which these examples do not test.

\FloatBarrier

\FloatBarrier

\section{Discussion and conclusions}
\label{sec: conclusions}

We have developed a statistical framework that predicts and fits optically thin synchrotron emission with Faraday rotation. 
The optically thin approximation is generally valid above the synchrotron self-absorption frequency, whereas at sufficiently low frequencies synchrotron self-absorption becomes important and the approximation breaks down. 
The statistical framework we developed represents each synchrotron emission spectrum as a linear combination of basis spectra derived from single-electron radiation, weighted by generalised moments of the electron population, magnetic field and Faraday depth (Fig.~\ref{fig: framework overview}). We compute the basis spectra once for all populations, and each prediction then costs the moments of the population and one matrix--vector product, which accelerates the calculations.

\paragraph{Population models and correlations.}
The statistical framework developed here accommodates general distributions of synchrotron-emitting electron populations. The generalised moments summarise the statistics of the emitting population. Joint moments encode correlations between electron energy, magnetic-field properties and Faraday depth. Specified distribution models are then applied as constraints on these moments. With the radiation physics, channel responses and expansion choices fixed, SEDs from different populations can be synthesised linearly using the same spectral basis.

We use the framework to predict intensity and linear polarisation from specified electron populations. The tests cover power-law, curved and aged electron spectra, for which second-order expansions within energy cells achieve sub-per-cent accuracy (Table~\ref{tab: classical fit summary}). Increasing the expansion order further improves the predictions. We also test populations in which the electron energy is correlated with magnetic-field strength or direction. Retaining these correlations through joint moments reduces the prediction errors by one to two orders of magnitude compared with calculations that neglect them.

\paragraph{Flexible SED fitting.}
Free-moment fits are linear and need no model of the electron spectrum. Every fit performed over bands up to $100\,\mathrm{GHz}$ matches the data to within the noise in intensity and linear polarisation (Fig.~\ref{fig: classical fits}), and they fit an independently computed \textsc{GALPROP} intensity spectrum well.

\paragraph{Extrapolation to higher frequencies.}
The spectral shape in the fitted band does not by itself fix the departure from a power law at higher frequencies. Electrons that emit mainly above the fitted band contribute to it through the low-frequency tails of their spectra, whose shapes vary little with electron energy. The band therefore measures their summed contribution but little of their energy distribution. Wider frequency coverage and lower noise constrain more of this distribution, and an electron-spectrum model fixes it through a few parameters.
Free-moment fits assume no electron-spectrum family. A non-negative electron distribution can be imposed as an intrinsic constraint. With this constraint, broad frequency coverage and low noise, the fits predict the intensity and linear polarisation of the tested smooth spectra to within a few per cent and closely follow the cut-off of the aged spectrum (Fig.~\ref{fig: cmb extrapolation}).
A parametric family that contains the true electron spectrum also achieves per-cent accuracy with noisier data and remains within a few per cent when fitted over a narrower band, provided the data constrain its parameters. A mismatched electron family can produce poor fits and extrapolations, as illustrated by the tested power-law-mixture electron energy models applied to the aged electrons. A wrong parametric form is not always visible in the fitted band. Fitted over $1$--$40\,\mathrm{GHz}$ at $1\%$ noise, the power-law mixtures fit the broken power law in most realisations and still miss its $300\,\mathrm{GHz}$ intensity by $20\%$ to $78\%$.
In the continuum limit, magnetic field-strength variations can be absorbed into an effective electron energy distribution under the stated independence assumptions. This effective distribution must lie within the adopted energy range and be representable by the fitted model.

\paragraph{Empirical spectral laws fail to fit or extrapolate emissions of complex electron spectra.}
At per-cent noise the best-matching empirical spectral law fits each intensity spectrum over $0.1$--$10\,\mathrm{GHz}$ to within the noise, while the curved power law leaves rms errors of $1.2\%$ for the \textsc{GALPROP} spectrum and $5.7$--$7.3\%$ for the aged spectra. As the noise falls the free-moment fits keep improving, whereas an empirical law stops improving once the spectrum departs from its functional form. At the lowest tested noise level, the free-moment fits are more accurate than the best tested empirical law for every spectrum that a curved power law does not describe, by factors of 10 to 40 for the broken power-law, aged and \textsc{GALPROP} spectra and by a factor of 3 for the two-zone spectrum (Table~\ref{tab: classical baselines}).
Extrapolation in either direction, upward to $300\,\mathrm{GHz}$ or downward to $1\,\mathrm{GHz}$, likewise depends on whether the empirical spectral law matches the true spectrum outside the fitted band (Table~\ref{tab: law extrapolation}). For the broken power law, empirical laws with a fixed two- or three-parameter form can pass the in-band $\chi^2$ test at per-cent noise yet miss the extrapolated intensity substantially. The four-parameter laws (the log-cubic law and the second-order $\beta$-moment SED with $\omega_1$ free) reduce this error, but still fail $\chi^2$ once the noise is low. For the aged spectrum, none of the tested empirical laws fit any of these bands.
%

\paragraph{Physical interpretation and parameter inference.}
A specified physical model predicts the moments. Conversely, an SED fit constrains combinations of the moments, which can then be used to infer physical parameters within a specified model. 
In tests with matching electron and field models, the fits recover the input parameters with uncertainties consistent with the scatter over noise realisations (Table~\ref{tab: classical parameters}). 
Comparisons between fits can also test a population assumption when the remaining model assumptions hold. In the two-zone example, the polarisation data reject statistical independence between magnetic-field azimuth and electron energy.
The field-strength and ageing tests also illustrate model-dependent parameter degeneracies. In the field-strength test, the spectral amplitude that combines electron density and field strength is recovered to better than 0.1\%, while the inferred density shifts by a few per cent. Adopting a different ageing model gives a similarly good fit but shifts the inferred age by more than ten per cent. Independent information on field strengths and pitch-angle evolution can help constrain densities and ages, with the uncertainties in that information propagated to the inferred quantities.

\paragraph{Extensions of the radiation model.}
Three extensions of the radiation model are reserved for future work. Keeping higher-order terms in $1/\gamma$ in the continuum kernels would allow them to describe circular polarisation, as the harmonic kernels already do.
For our test populations with isotropic pitch angles, we estimate that the corrections to $I$ and $Q$ are below $10^{-6}$ of the intensity over $0.1$--$300\,\mathrm{GHz}$ (Sec.~\ref{sec: continuum limit}).
Steady bulk motion shifts the observed frequencies and changes both the brightness and the angle from which we see the magnetic field. For a given distribution of bulk velocities, the moments of the electrons at each velocity transform linearly, and their average over the velocities gives the population moments, with the same single-electron spectra throughout (Sec.~\ref{sec: local population}).
At low frequencies, the local synchrotron self-absorption coefficient is also a linear sum of cell moments~\cite{rybicki1979radiative,pandya2016symphony}. Each region can absorb radiation from those behind it, so the observed spectrum depends on their order along the line of sight and is no longer linear in the moments (Sec.~\ref{sec: screen model}).

\vspace{1em}

The numerical tests cover intensity and linear polarisation along one line of sight with an independent Faraday screen, in the ultra-relativistic continuum limit, where the harmonic kernels reproduce the classical synchrotron results. The formalism also covers the harmonic regime at lower electron energies, circular polarisation and correlations between emission and Faraday depth. Other next steps are population tests of these cases, joint fits of the field and pitch-angle distributions with the electron spectrum, Faraday conversion, and the telescope response with its beam and sky averaging for applications to observed maps. \textsc{SyncEM} implements the basis construction and fitting procedures of this paper and provides the starting point for these extensions for radio and CMB foreground modelling.

\FloatBarrier

\appendix

\section{From harmonic amplitudes to the continuum kernels}
\label{sec: continuum derivation}

We reduce the harmonic amplitudes of Eq.~\eqref{eq: harmonic amplitudes} to the kernels of Eq.~\eqref{eq: directional continuum}, retaining the source-time normalisation. Take $\gamma\gg1$ at fixed $x=\nu/\nu_{\rm c}>0$ and fixed $\theta$ away from the field axis. The relevant pitch angles have $y=\gamma(\alpha-\theta)=O(1)$. Put
\begin{equation}
 \varepsilon=1-\left(\frac{z_m}{m}\right)^2,\qquad
 \chi(y)=\frac{x}{2}(1+y^2)^{3/2}.
 \label{eq: continuum angular variables}
\end{equation}
With $z_m$ of Eq.~\eqref{eq: harmonic amplitudes} the first relation below is exact. Its leading order in $1/\gamma$ gives the second, and the line frequency $\nu=\nu_m$ then gives the third
\begin{equation}
 D^2\varepsilon=(\cos\theta-\beta\cos\alpha)^2+\gamma^{-2}\sin^2\theta,\quad
 \varepsilon\simeq\frac{1+y^2}{\gamma^2\sin^2\theta},\quad
 \frac{m\varepsilon^{3/2}}3\simeq\chi(y).
 \label{eq: continuum angular geometry}
\end{equation}
The large-order turning-point expansion, written using the Airy--Bessel relations~\cite[\S\S9.6(i), 10.19(iii)]{nistDLMF}, gives
\begin{equation}
 J_m(z_m)\simeq\frac{\sqrt\varepsilon}{\pi\sqrt3}K_{1/3}(\chi),\qquad
 J_m'(z_m)\simeq\frac{\varepsilon}{\pi\sqrt3}K_{2/3}(\chi).
 \label{eq: harmonic bessel continuum}
\end{equation}
Replacing the harmonic sum by its frequency integral introduces the line density $\diff m/\diff\nu$. With $\nu_B=\omega_B/(2\pi)$ and $\omega=2\pi\nu$, the source-time prefactor and the pitch measure give
\begin{equation}
 \frac{\diff m}{\diff\nu}=\frac{D}{\nu_B},\qquad
 C_m^2\frac{\diff m}{\diff\nu}=\frac{e^2\omega^2}{c\omega_B},\qquad
 |\diff\mu|\simeq\frac{\sin\theta}{\gamma}\diff y.
 \label{eq: continuum normalisation reduction}
\end{equation}
A pitch-angle density that varies little over $y=O(1)$ is evaluated at $\mu=\eta$ and taken outside the pitch integral, as in Eq.~\eqref{eq: beaming limit}. The continuum kernels $\tfrac12\int_{-1}^{1}\mathcal K^{\rm nat}\diff\mu$ of the two linear-component powers of Eq.~\eqref{eq: harmonic prefactor} are then, at leading order,
\begin{equation}
 \begin{pmatrix}
 \mathcal K_\perp^{\rm cont}\\
 \mathcal K_\parallel^{\rm cont}
 \end{pmatrix}
 =\frac{\sqrt3A_\theta x^2}{2\pi}
 \int_{-\infty}^{\infty}
 \begin{pmatrix}
 (1+y^2)^2K_{2/3}^2(\chi)\\
 y^2(1+y^2)K_{1/3}^2(\chi)
 \end{pmatrix}\diff y.
 \label{eq: continuum polarised angular integrals}
\end{equation}
For $x>0$, these angular integrals evaluate to~\cite{westfold1959polarization}
\begin{equation}
 \begin{aligned}
 \int_{-\infty}^{\infty}(1+y^2)^2K_{2/3}^2(\chi)\diff y
 &=\frac{\pi}{\sqrt3x^2}[F(x)+G(x)],\\
 \int_{-\infty}^{\infty}y^2(1+y^2)K_{1/3}^2(\chi)\diff y
 &=\frac{\pi}{\sqrt3x^2}[F(x)-G(x)],
 \end{aligned}
 \label{eq: continuum angular identities}
\end{equation}
where $F,G$ are the functions in Eq.~\eqref{eq: directional continuum}. Hence
\begin{equation}
 \mathcal K_\perp^{\rm cont}=\frac{A_\theta}{2}(F+G),\qquad
 \mathcal K_\parallel^{\rm cont}=\frac{A_\theta}{2}(F-G).
 \label{eq: continuum linear components}
\end{equation}
The Stokes parameters $I=\parallel+\perp$ and $Q=\parallel-\perp$ give $\mathcal K_I^{\rm cont}=A_\theta F$ and $\mathcal K_Q^{\rm cont}=-A_\theta G$.
For $V_m$ of Eq.~\eqref{eq: Vn}, the corresponding integrand is $2y(1+y^2)^{3/2}K_{1/3}(\chi)K_{2/3}(\chi)$. It is odd in $y$, so the integral vanishes at this order, and we neglect $\mathcal K_V^{\rm cont}$.

Figure~\ref{fig: harmonic basis spectra} shows basis spectra at $\gamma_0=100$ computed in \textsc{SyncEM} by angular quadrature of the harmonic kernels, without the continuum approximation and without a bound on the numerical error. Doubling the angular nodes changes the $I$ and $Q'$ spectra by less than $10^{-4}$ of their peaks and the $V$ spectra by up to $2\times10^{-3}$. After one common rescaling that matches the peaks of the $(0,0)$ intensity, the $I$ and $Q'$ spectra agree with the basis spectra from the continuum kernels to better than $10^{-3}$ of their peaks, and their field derivatives follow from the energy derivatives (Appendix~\ref{app: test ensembles}) to the same accuracy. The finite beaming cone adds $V$ on the pairs with $\ell+k$ odd, at up to a few per cent of the intensity.

\section{Test ensembles and comparisons}
\label{app: test ensembles}

\paragraph{Ensembles.}
Table~\ref{tab: classical ensembles} lists the ensembles of Sec.~\ref{sec: classical models} and the settings they share. E1 and E3 follow Eq.~\eqref{eq: index mixture}, and the E2 and E4 electron densities are
\begin{align}
 \text{E2:}&\quad \mathcal N\propto(\gamma/\gamma_{\rm br})^{-p_1}\left[1+(\gamma/\gamma_{\rm br})^s\right]^{-(p_2-p_1)/s},\nonumber\\
 \text{E4:}&\quad \mathcal N\propto\gamma^{-p}(1-b\gamma t)^{p-2},\quad \gamma<\gamma_c=1/(bt).
 \label{eq: reference densities}
\end{align}
In E4, $t$ is the time since injection, $b=b_s+b_c$ for JP and $b(\mu)=\tfrac32b_s(1-\mu^2)+b_c$ for KP. Here $b_s\gamma^2$ is the pitch-averaged synchrotron loss rate $|\diff\gamma/\diff t|$ at $\langle B^2\rangle_{\rm col}=25.3\,\mu\mathrm G^2$, and $b_c\gamma^2$ is the Thomson inverse-Compton loss rate in a radiation field of energy density $1\,\mathrm{eV\,cm^{-3}}$. The aged form assumes an impulsive power-law injection without an upper cut-off. We use it on $[100,\gamma_c(\mu))$.

\begin{table}[tbp]
\centering
\small
\caption{Test ensembles. The upper block gives the settings shared by all ensembles unless a row states otherwise. The last column gives the origin of each model. The parameter values are our choice.}
\label{tab: classical ensembles}
\begin{tabular}{@{}llll@{}}
\toprule
\multicolumn{2}{@{}l}{Shared setting} & \multicolumn{2}{l@{}}{Default} \\
\midrule
\multicolumn{2}{@{}l}{electron energy} & \multicolumn{2}{l@{}}{$\gamma\in[100,2.96\times10^{6}]$} \\
\multicolumn{2}{@{}l}{field strength} & \multicolumn{2}{l@{}}{$B\sim U[4,6]\,\mu$G} \\
\multicolumn{2}{@{}l}{field inclination} & \multicolumn{2}{l@{}}{$\eta=\cos\theta\sim U[-1/2,1/2]$} \\
\multicolumn{2}{@{}l}{pitch angles} & \multicolumn{2}{l@{}}{isotropic} \\
\multicolumn{2}{@{}l}{projected-field azimuth} & \multicolumn{2}{l@{}}{$\langle e^{2i\phi}\rangle_{\rm F}=0.4e^{0.5i}$} \\
\multicolumn{2}{@{}l}{Faraday screen} & \multicolumn{2}{l@{}}{independent Gaussian, mean $8$, s.d.\ $0.5\,\mathrm{rad\,m^{-2}}$} \\
\midrule
Name & Model & Parameters & Ref. \\
\midrule
E1 & power law & $p=2.6$ & \cite{ginzburg1965cosmic,rybicki1979radiative} \\
E2 & smoothly broken power law & $p_1=2.0$, $p_2=3.2$, $s=2$, $\gamma_{\rm br}=7800$ & \cite{strong2011electron,orlando2018leptons} \\
E3 & Gaussian index mixture & $\bar p=2.6$, $\sigma_p=0.2$, $\gamma_{\rm piv}=10^{4}$ & \cite{chluba2017rethinking} \\
E4-JP & ageing, isotropised pitch & $p=2.6$, $t=19$\,Myr, $\gamma_c=19673$ & \cite{jaffe1973dynamical} \\
E4-KP & ageing, fixed pitch & as E4-JP, with $\gamma_c(\mu)=16489$--$32049$ & \cite{kardashev1962nonstationarity,pacholczyk1970radio} \\
E5 & two power-law zones & $p_A=3.1$, $p_B=2.1$, $\Delta\phi=45^\circ$, $f_A=0.938$ & \cite{vacher2023spin} \\
E6 & ageing, local field & as E4-JP, losses at local $B$, $\gamma_c(B)=16950$--$22900$ & \cite{tribble1993radio} \\
E1M & E1 in a Maxwellian field & $B_{\rm rms}=5\,\mu$G & \cite{tribble1993radio} \\
\bottomrule
\end{tabular}
\end{table}

E2 steepens near $4\,$GeV, as do \textsc{GALPROP}-propagated local electron spectra~\cite{strong2011electron,orlando2018leptons}. The E4-JP break frequency is $\nu_b=3e\langle B_\perp\rangle_{\rm col}\gamma_c^2/(4\pi m_ec)=7.77\,$GHz. E5 places two power-law zones behind a common screen, with azimuth densities $[1+0.8\cos2(\phi-\phi_z)]/(2\pi)$, $\phi_A=0.25\,$rad and $\phi_B=\phi_A+\pi/4$. The zones have equal intensity at $1\,$GHz, and $f_A$ is the electron fraction in zone A. Because the zones differ in spectrum and polarisation angle, the polarisation angle of their sum depends on frequency~\cite{vacher2023spin}. E6 evaluates the synchrotron loss of E4-JP in the local field~\cite{tribble1993radio}. E1M places the E1 electrons in fields with strength density $\sqrt{2/\pi}\,B^2\sigma_B^{-3}e^{-B^2/(2\sigma_B^2)}$, $\sigma_B=B_{\rm rms}/\sqrt3$, which is the strength distribution of an isotropic Gaussian random field. Its inclination and azimuth statistics are the defaults of Table~\ref{tab: classical ensembles}.

\begin{figure}[tbp]
\centering
\includegraphics[width=\textwidth]{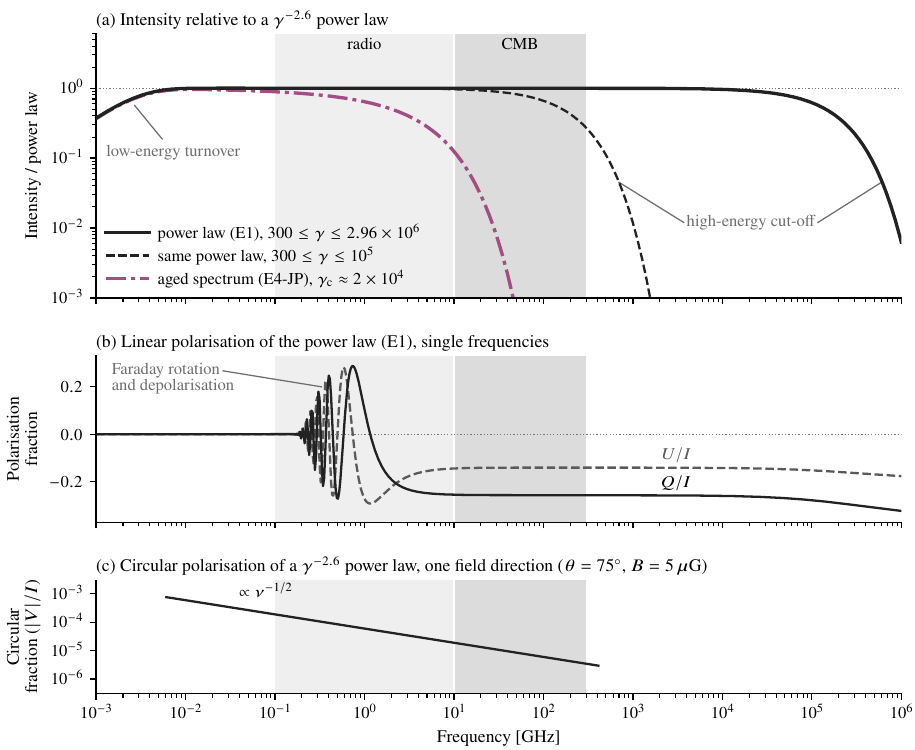}
\caption{Intensity, linear polarisation and circular polarisation of the ensembles at single frequencies. Shading marks the radio band ($0.1$--$10\,$GHz) of the main tests and the CMB band ($10$--$300\,$GHz). (c)~$|V|/I$ for one field direction, from the fitting formula of Ref.~\cite{pandya2016symphony}.}
\label{fig: emission windows}
\end{figure}

Figure~\ref{fig: emission windows} shows the energy cut-offs, ageing and Faraday screen of the ensembles. With $\gamma\leq2.96\times10^6$, the intensity of a $\gamma^{-2.6}$ spectrum stays within $10^{-3}$ of the untruncated power law up to $300\,$GHz. An upper limit of $10^5$ instead leaves it $3.4\%$ low at $10\,$GHz.

\paragraph{Basis and fit settings.}
The basis uses 41 channels with centres $\nu_j$ logarithmically spaced over $0.1$--$10\,\mathrm{GHz}$. Their channel responses are the unit-area bumps
\begin{equation}
    R_j(\nu)=\frac{\rho[(\nu-\nu_j)/h_j]}{h_jZ},\qquad
    \rho(\xi)=\exp\!\left(-\frac{\xi^2}{1-\xi^2}\right)\ \ (|\xi|<1),\qquad
    h_j=\frac{f\nu_j}{\sigma_\rho},
    \label{eq: forward response families}
\end{equation}
with $\rho=0$ for $|\xi|\geq1$, $Z=\int_{-1}^{1}\rho\diff\xi$ and $\sigma_\rho^2=Z^{-1}\int_{-1}^{1}\xi^2\rho\diff\xi$. With $\sigma_\rho\approx0.40$, this choice of $h_j$ makes $f\nu_j$ the rms width of $R_j$ about $\nu_j$, so $f$ is the fractional rms width of Sec.~\ref{sec: channel response}. The tests use $h_j=0.2\nu_j$ ($f=8.0\%$).

The tests specify the conditional distribution of field strength, angles, azimuth and Faraday depth at each energy and fit only the energy distribution. These variables are averaged into the channel kernel, giving $\overline{\mathcal K}_{Xj}(\gamma)$ for $X\in\{I,V,P\}$, including the azimuthal factor $e^{2i\phi}$ and the Faraday phase for $X=P$. If that conditional distribution depends on $\gamma$, the averaged kernel carries this dependence. The energy range is divided into cells of equal width in $\ln\gamma$, the lattice of Sec.~\ref{sec: cell lattice} along $\ln\gamma$ with the other variables integrated out. For cell $b$, let $\gamma_b$ be its geometric centre and $h_b$ its half-width in $\ln\gamma$, and define
\begin{equation}
    u_b=\frac{\ln(\gamma/\gamma_b)}{h_b}\in[-1,1],\qquad
    \widetilde X_{jbr}
    =\left.\frac{1}{r!}\partial_{u_b}^{\,r}
    \overline{\mathcal K}_{Xj}(\gamma_b e^{h_bu_b})\right|_{u_b=0},
    \quad r=0,\dots,N.
    \label{eq: illustrated derivative spectra}
\end{equation}
The moments of cell $b$ are the integrals of $u_b^r$ over the electron column in the cell. For the linear fit of Sec.~\ref{sec: linear fit}, each moment is divided by a fixed positive reference weight $w_b$ and its basis function is multiplied by $w_b$, which leaves the prediction unchanged. The basis functions $\widetilde X_{jbr}$ contain the specified conditional statistics, and the moments carry the unknown energy distribution. The joint energy--field basis used for E6 (Sec.~\ref{sec: classical ensembles}) also needs field derivatives of the basis functions. For the continuum kernels they follow from the energy derivatives when the distributions averaged into $\overline{\mathcal K}_{Xj}$ are held fixed and do not depend on energy, because the field strength then enters Eq.~\eqref{eq: directional continuum} only through $A_\theta\propto B_\perp$ and $\nu_{\rm c}\propto B_\perp\gamma^2$.

The energy range $300\leq\gamma\leq\gamma_{\rm max}$, with $\gamma_{\rm max}\simeq2.96\times10^6$, is divided into nineteen cells of equal width in $\ln\gamma$, with half-width $h_b=\ln(\gamma_{\rm max}/300)/38\simeq0.242$. With $N=2$ each cell has three energy moments, 57 in total, and each moment multiplies one column of $I$, $Q$ and $U$ channel values. Two further cells cover $100\leq\gamma<300$ with the harmonic kernels. The amplitude of each spectrum is set so that the electron column at $\gamma\geq300$ is one, and the fraction $f_A$ of E5 refers to these energies. The columns of these two cells are below $10^{-10}$ of the others in the channels used, so the ranks and fits quoted below are those of the 57 moments. In \textsc{SyncEM} the continuum columns are computed on 512 cells in $\ln\nu_{\rm c}$ at order eight and projected onto these energy cells. The reference weights are
\begin{equation}
    w_b=\int_{|u_b|\leq1}n_{\rm ref}(\gamma)\diff\gamma,\qquad
    n_{\rm ref}(\gamma)=\frac{1.6\gamma^{-2.6}}{300^{-1.6}-\gamma_{\rm max}^{-1.6}}.
    \label{eq: fit coefficient scales}
\end{equation}
The power law $n_{\rm ref}(\gamma)$ in these weights does not force the fitted moments to follow a power law, but the unconstrained combinations and the extrapolation of the free-moment fits depend on this choice.

For E5 the 57 energy moments and the complex azimuth moments of each cell give 171 real columns. For E6 the joint columns are the six monomials of total degree at most two in $u_b$ and $(B-5\,\mu\mathrm G)/(1\,\mu\mathrm G)$ in each cell. The rank search runs up to the numerical rank of the noise-weighted basis matrix, which is 26 for the 57 energy columns and 61 for E5. The intensity-only fits of Table~\ref{tab: classical baselines} use 21 intensities, and their rank search stops at $R=14$. Removing the limit does not change which method is more accurate in any column of that table. Doubling the quadrature nodes of the true spectra changes them by at most $3\times10^{-14}I_j$.

In E4-KP the pitch-angle density depends on energy, so the factorisation of Eq.~\eqref{eq: uncoupled moments} does not apply. Legendre columns in the pitch angle up to $\ell=8$ include this coupling. They are evaluated at $\mu=\eta$, where the continuum kernels place the emission. From the moments of the KP distribution they reproduce the spectrum to $0.88\%$ at $N=2$ and $0.066\%$ at $N=4$. A JP spectrum with $t_{\rm eff}=1.1475\,t$ reproduces the E4-KP spectrum to $7.7\times10^{-5}I_j$, and the energy basis at its moments reproduces the E4-KP spectrum to $0.89\%$ at $N=2$. Table~\ref{tab: classical fit summary} uses these JP moments as the E4-KP reference.

For the parameter fits of Sec.~\ref{sec: moment combinations}, over 1000 noise realisations the one-standard-deviation intervals cover the truth in $65.7$--$69.1\%$ of the fits for E1--E4-JP, and the $N=2$ representation error shifts them by at most $0.26$ standard deviations. Parameter uncertainties from the compressed and full linear likelihoods differ by at most $0.3\%$. For E1M the ratio $0.964$ of $\langle B_\perp^{1.8}\rangle_{\rm col}$ in the Maxwellian field to that for $B\sim U[4,6]\,\mu$G accounts for the shift of the density amplitude. Over 1000 noise realisations the JP fit to E4-KP gives $t_{\rm eff}=(1.150\pm0.006)\,t$, where $t$ is the true age.

\paragraph{Comparison laws.}
Table~\ref{tab: classical baselines} compares the basis with empirical spectral laws on the intensity alone, because \textsc{PySM} applies the same law to $Q$ and $U$ with a fixed polarisation angle. With $y=\ln(\nu/1\,\mathrm{GHz})$, the laws are the power law $I_0e^{\beta y}$ and curved power law $I_0e^{\beta y+Cy^2}$ of the \textsc{PySM} models s1 and s3~\cite{thorne2017pysm,zonca2021pysm3}, the log-cubic law $\exp(c_0+c_1y+c_2y^2+c_3y^3)$, and the second-order $\beta$-moment SED $I_0e^{\bar\beta y}(1+\omega_1y+\omega_2y^2/2)$ of Ref.~\cite{chluba2017rethinking}. All are integrated through the channel responses. We fit $I_0$, $\beta$ and $C$ to each spectrum instead of using the \textsc{PySM} templates. Here $\beta$ is the intensity index, which exceeds the brightness-temperature index of \textsc{PySM} by 2. Ref.~\cite{chluba2017rethinking} chooses $\bar\beta$ so that $\omega_1=0$, and we fit both this form and the form with $\omega_1$ free. Neither fit constrains the parameters to be moments of a positive distribution.

The electron-spectrum families are fitted through the same 57 columns. They are the cumulant form of Eq.~\eqref{eq: index cumulant expansion} with $\gamma_{\rm piv}=10^4$ truncated at first order (ePL) and at second order with $k_2$ of either sign (eLP), and the E2 form with $s=2$ (eSBPL). E1 is a member of ePL and eLP, and E3 of eLP. The fits bound $k_1$ to $[0.5,6]$, $k_2$ to $[-1,1]$, $p_1$ and $p_2$ to $[0,7]$, and $\gamma_{\rm br}$ to $[100,3\times10^5]$.

The \textsc{GALPROP} spectrum comes from the intensity map z10LMPD\_\allowbreak SUNfE, which corresponds to the SUN10E model of Ref.~\cite{orlando2013synchrotron} (their Fig.~9).\footnote{\url{https://www.mpe.mpg.de/~aws/test_healpix_standard_synchrotron_healpix_55_z10LMPD_SUNfE.gz}} Its full setup is not public (for example, whether it includes free-free absorption), so we test only how well the basis fits this spectrum. We average it over $|b|\geq60^\circ$ at its native frequencies $10^{-3}\times1.2^{26+k}\,$GHz with $k=0,\dots,24$ ($0.114$--$9.10\,$GHz). We fit even $k$ and withhold odd $k$. The basis is evaluated at these frequencies without a channel response, which the continuum kernels allow because they contain no harmonic lines.

\begin{table}[tbp]
\centering
\footnotesize
\setlength{\tabcolsep}{2pt}
\caption{Rms error of intensity-only fits in the channels left out of the fit, in per cent of $I_j$: medians over 500 realisations with $0.01\%/1\%$ noise. Par.\ gives the parameter count or the range of median UPRE ranks across ensembles and noise levels. Free-moment fits use $N=2$, with $R=4$ on noise-free data or $R\leq14$ selected by UPRE. The electron families also use the $N=2$ basis, whose representation error limits their low-noise accuracy even for the family that generated the data. With $\omega_1$ free, $\bar\beta$ and $\omega_1$ are nearly degenerate, so that SED acts as a generic four-parameter law.}
\label{tab: classical baselines}
\begin{tabular}{@{}lcccccccc@{}}
\toprule
Model & Par. & E1 & E2 & E3 & E4-JP & E4-KP & E5 & \textsc{GALPROP} \\
\midrule
free moments, $R=4$ & 4 & 4.1 & 2.4 & 4.4 & 5.1 & 6.3 & 5.5 & 2.8 \\
free moments, UPRE & 7--14 & 0.0089/0.65 & 0.0080/0.62 & 0.0092/0.66 & 0.012/0.69 & 0.014/0.70 & 0.011/0.68 & 0.033/0.78 \\
PySM s1 form & 2 & 0.0026/0.26 & 8.0/8.0 & 0.80/0.84 & 21/21 & 26/26 & 4.6/4.6 & 8.0/8.0 \\
PySM s3 form & 3 & 0.0033/0.32 & 0.22/0.39 & 0.0033/0.32 & 5.7/5.7 & 7.3/7.3 & 0.064/0.34 & 1.2/1.2 \\
$\beta$-mom., $\omega_1=0$ & 3 & 0.0033/0.32 & 0.47/0.58 & 0.0043/0.33 & 5.5/5.5 & 7.0/7.0 & 0.034/0.33 & 1.3/1.4 \\
log-cubic & 4 & 0.0039/0.39 & 0.19/0.42 & 0.0039/0.39 & 1.4/1.4 & 1.8/1.8 & 0.064/0.39 & 0.33/0.57 \\
$\beta$-mom., $\omega_1$ free & 4 & 0.0039/0.39 & 0.32/0.55 & 0.0047/0.39 & 0.51/0.62 & 0.46/0.58 & 0.033/0.39 & 0.41/0.63 \\
electron family ePL & 2 & 0.020/0.26 & 8.0/8.0 & 0.80/0.84 & 21/21 & 26/26 & 4.6/4.7 & 8.0/8.0 \\
electron family eLP & 3 & 0.020/0.33 & 0.20/0.38 & 0.021/0.33 & 6.2/6.2 & 10/10 & 0.070/0.34 & 1.2/1.3 \\
electron family eSBPL & 4 & 0.021/0.35 & 0.020/0.39 & 0.032/0.38 & 3.1/3.1 & 4.7/4.7 & 0.094/0.40 & 0.27/0.54 \\
\bottomrule
\end{tabular}
\end{table}

\paragraph{Extrapolation tests.}
The extrapolation tests of Sec.~\ref{sec: cmb extrapolation} use the channel responses above with $h_j=0.2\nu_j$, centred on a logarithmic grid from $1\,\mathrm{GHz}$ with a ratio of $40^{1/20}$ between neighbours. The fits use the first 21 channels ($1$--$40\,\mathrm{GHz}$) or the first 26 ($1$--$100.6\,\mathrm{GHz}$). The downward test of Table~\ref{tab: law extrapolation} fits channels 12--32 ($7.6$--$304\,\mathrm{GHz}$) with the same cells and predicts the $1\,\mathrm{GHz}$ channel. The predictions are evaluated in a separate set of 121 channels with the same channel responses, logarithmically spaced over $1$--$300\,\mathrm{GHz}$. The noise is independent and Gaussian, with standard deviation $0.01I_j$ or $10^{-4}I_j$ in each of $I$, $Q$ and $U$, and each test uses 500 realisations shared by all methods. Free moments and non-negative cells use the cells above $\gamma=1.28\times10^3$ (cells 4--19), and the electron families use all cells. Non-negative cells place 17 point masses at Chebyshev--Lobatto points in each cell and fit their weights by non-negative least squares with a penalty $\alpha|\boldsymbol a|^2$ on the cell moments, where $\alpha$ minimises the UPRE of the same fit without positivity. The moment and cumulant forms use $\gamma_{\rm piv}=10^4$. The moment form is linear in $Am_n$ (Eq.~\eqref{eq: index moment expansion}). The cumulant form is the eLP family of Table~\ref{tab: classical baselines}, with the same bounds, and is fitted with the full linear likelihood of Sec.~\ref{sec: linear fit}. The eSBPL fits to E2 use the bounds of that table and the same likelihood. The in-band goodness-of-fit test accepts a fit when the probability of a larger $\chi^2$ is at least $0.001$. The degrees of freedom are the number of fitted data values minus the number of fitted parameters, or minus the effective number of parameters for non-negative cells. Figure~\ref{fig: cmb extrapolation} shows the data of the first realisation.
The bound on the $300\,\mathrm{GHz}$ intensity of E4-JP in Sec.~\ref{sec: cmb extrapolation} is the largest value among non-negative cell fits without the penalty whose $\chi^2$ exceeds their minimum by at most one. Over the first 100 realisations at $1\%$ noise its median is $1.7\times10^{-5}$ of the power law for fits over $1$--$100\,\mathrm{GHz}$ and $0.21$ for fits over $1$--$40\,\mathrm{GHz}$. The intervals in Fig.~\ref{fig: cmb extrapolation} come from the noise alone and exclude uncertainty in the field, screen and electron-family assumptions.

\paragraph{Extrapolation with empirical laws.}
Table~\ref{tab: law extrapolation} gives the errors. For E1 and E3 the curved power law predicts the target intensity with median errors below $2\%$ at $1\%$ noise in all three bands. For E2, at both noise levels, it misses the $300\,\mathrm{GHz}$ intensity by $19\%$ when fitted over $1$--$40\,\mathrm{GHz}$ and the $1\,\mathrm{GHz}$ intensity by $23\%$ when fitted over $7.6$--$304\,\mathrm{GHz}$, although at $1\%$ noise its in-band $\chi^2$ accepts $98\%$ of the $1$--$40\,\mathrm{GHz}$ fits. Free moments fitted over $7.6$--$304\,\mathrm{GHz}$ are accepted for E1--E3 at both noise levels and miss the $1\,\mathrm{GHz}$ intensity by $55$--$69\%$. The in-band $\chi^2$ rejects every law fitted to E4-JP in these bands, and the free-moment fits to E4-JP over $7.6$--$304\,\mathrm{GHz}$.

\begin{table}[tbp]
\centering
\footnotesize
\setlength{\tabcolsep}{3pt}
\caption{Error of the predicted intensity at the target frequency, in per cent of the true intensity: medians over 500 realisations with $0.01\%/1\%$ noise. Bands and targets are in GHz. The laws are those of Table~\ref{tab: classical baselines}, fitted to $I$, with the pivot of the $\beta$-moment SED with $\omega_1=0$ at the geometric centre of each band. Free moments are fitted to $I$, $Q$ and $U$ as in Sec.~\ref{sec: cmb extrapolation}. A $^\times$ marks errors of fits that the in-band $\chi^2$ rejects in most realisations. E4-JP is omitted because the in-band $\chi^2$ rejects every law fitted to it.}
\label{tab: law extrapolation}
\begin{tabular}{@{}llcccccc@{}}
\toprule
Ensemble & Band $\to$ target & PySM s1 & PySM s3 & $\beta$-mom., $\omega_1=0$ & log-cubic & $\beta$-mom., $\omega_1$ free & free moments \\
\midrule
E1 & $1$--$40\to300$ & 0.070/0.56 & 0.058/1.9 & 0.058/1.9 & 0.076/6.9 & 0.067/5.8 & 14/62 \\
 & $1$--$100\to300$ & 0.062/0.37 & 0.046/0.82 & 0.046/0.82 & 0.031/2.0 & 0.033/1.9 & 2.1/24 \\
 & $7.6$--$304\to1$ & 0.048/0.54 & 0.075/1.9 & 0.075/1.9 & 0.096/7.4 & 0.12/6.6 & 57/65 \\
\addlinespace
E2 & $1$--$40\to300$ & 34$^\times$/34$^\times$ & 19$^\times$/19 & 33$^\times$/33 & 2.2$^\times$/6.8 & 6.2$^\times$/9.2 & 19/45 \\
 & $1$--$100\to300$ & 20$^\times$/20$^\times$ & 11$^\times$/11 & 15$^\times$/15 & 2.0$^\times$/2.7 & 2.9$^\times$/3.0 & 1.6/16 \\
 & $7.6$--$304\to1$ & 43$^\times$/43$^\times$ & 23$^\times$/23 & 21$^\times$/21 & 6.4$^\times$/9.0 & 18$^\times$/16 & 58/69 \\
\addlinespace
E3 & $1$--$40\to300$ & 6.6$^\times$/6.6 & 0.10/1.9 & 0.11/1.8 & 0.086/6.9 & 0.15/5.9 & 17/65 \\
 & $1$--$100\to300$ & 4.7$^\times$/4.6 & 0.078/0.83 & 0.019/0.78 & 0.049/2.0 & 0.048/1.9 & 2.3/26 \\
 & $7.6$--$304\to1$ & 6.7$^\times$/6.7 & 0.13/1.9 & 0.34/1.8 & 0.14/7.3 & 0.091/6.7 & 55/64 \\
\bottomrule
\end{tabular}
\end{table}

\begin{table}[tbp]
\centering
\small
\setlength{\tabcolsep}{4pt}
\caption{Physical parameters fitted to the retained combinations at $R_\star$ (compressed, Eq.~\eqref{eq: compressed likelihood}) and with the full linear likelihood, for the noise realisation of Table~\ref{tab: classical fit summary}. $A$ is the electron density per unit $\ln\gamma$ at $\gamma=10^4$ for E1 and E3, and that of the power law without the cut-off for E4-JP. In each family, including E2 with $s=2$ fixed, the prefactor of $\diff\mathcal N/\diff\gamma$ is $A/10^4$.}
\label{tab: classical parameters}
\begin{tabular}{@{}llccc@{}}
\toprule
Ensemble ($R_\star$) & Parameter & Truth & Compressed & Full \\
\midrule
E1 (7) & $\ln A$ & $-5.1405$ & $-5.1425\pm0.0022$ & $-5.1425\pm0.0022$ \\
 & $p$ & $2.6000$ & $2.5977\pm0.0031$ & $2.5976\pm0.0031$ \\
\addlinespace
E2 (7) & $\ln A$ & $-2.97$ & $-2.63\pm0.36$ & $-2.62\pm0.36$ \\
 & $p_1$ & $2.000$ & $1.933\pm0.074$ & $1.930\pm0.074$ \\
 & $p_2$ & $3.200$ & $3.151\pm0.054$ & $3.149\pm0.054$ \\
 & $\ln\gamma_{\rm br}$ & $8.96$ & $8.83\pm0.14$ & $8.83\pm0.14$ \\
\addlinespace
E3 (7) & $\ln A$ & $-5.3161$ & $-5.3172\pm0.0046$ & $-5.3171\pm0.0046$ \\
 & $\bar p$ & $2.6000$ & $2.5980\pm0.0033$ & $2.5980\pm0.0033$ \\
 & $\sigma_p^2$ & $0.040$ & $0.0377\pm0.0096$ & $0.0376\pm0.0096$ \\
\addlinespace
E4-JP (9) & $\ln A$ & $-5.1159$ & $-5.1190\pm0.0053$ & $-5.1189\pm0.0053$ \\
 & $p$ & $2.6000$ & $2.5987\pm0.0063$ & $2.5986\pm0.0063$ \\
 & $\ln(\nu_b/{\rm Hz})$ & $22.774$ & $22.774\pm0.013$ & $22.773\pm0.013$ \\
\bottomrule
\end{tabular}
\end{table}

\FloatBarrier

\section*{Code availability}

The spectral basis and fits are implemented in \textsc{SyncEM} (\texttt{pip install syncem}), available from \href{https://github.com/zzhang0123/syncem}{\text{github.com/zzhang0123/syncem}} with online documentation \href{https://syncem.readthedocs.io}{\text{syncem.readthedocs.io}}. The numerical tests of Sec.~\ref{sec: classical models} and Appendix~\ref{app: test ensembles}, including Figs.~\ref{fig: harmonic continuum} and \ref{fig: harmonic basis spectra}, use version~\href{https://github.com/zzhang0123/syncem/releases/tag/v0.9.0}{0.9.0}, whose \texttt{studies/\allowbreak classical-models} holds the scripts.
The output data files and figures are in the release archive \texttt{classical-models-\allowbreak outputs-\allowbreak v0.9.0.tar.gz}.

\section*{Acknowledgements}
The results were obtained as part of the RadioForegroundsPlus Project HORIZON-CL4-2023-SPACE-01, GA 101135036 (JC; RW; ZZ).
We furthermore acknowledge support by the European Research Council (ERC) under the European Union's Horizon 2020 research and innovation programme (Grant agreement No. 948764; ZZ).
ZZ thanks OpenAI for providing free access through its researcher plan.
AI tools were used to implement and test code and to critically review and improve drafts. The authors checked the results and take responsibility for the content.
This work used \textsc{jax}~\cite{bradbury2018jax},
\textsc{equinox}~\cite{kidger2021equinox}, \textsc{numpy}~\cite{harris2020array},
\textsc{scipy}~\cite{virtanen2020scipy}, \textsc{matplotlib}~\cite{hunter2007matplotlib}
and \textsc{mpmath}~\cite{mpmath2023mpmath}.

\bibliographystyle{JHEP}
\bibliography{main}

\end{document}